\documentclass[showpacs,aps,floatfix,prb,reprint,superscriptaddress,10pt]{revtex4-1}

\usepackage{xfrac}
\usepackage{amssymb}
\usepackage{braket}
\usepackage{graphicx}
\usepackage{verbatim}
\usepackage{etoolbox}
\usepackage{amsfonts}
\usepackage{amsmath}
\usepackage[utf8]{inputenc}
\usepackage{mathtools}
\usepackage[dvipsnames]{xcolor}
\usepackage{soul,colortbl}
\usepackage{hyperref} \hypersetup{colorlinks=true,citecolor=blue,linkcolor=blue,urlcolor=black}
\usepackage{bm}
\usepackage{array}
\newcolumntype{C}[1]{>{\centering\arraybackslash}p{#1}}\usepackage{soul}
\definecolor{Gray}{gray}{0.85}

\usepackage{color}
\usepackage{morefloats}

\bibpunct{[}{]}{,}{n}{}{}

\definecolor{Gray}{gray}{0.9}
\definecolor{LightCyan}{rgb}{0.88,1,1}

\def\acoustic{{\mathrm{a}}}

\def\EOS{{\small EOS}}
\def\VRH{{\small VRH}}
\def\AGL{{\small AGL}}
\def\AEL{{\small AEL}}
\def\AFLOW{{\small AFLOW}}
\def\AFLOWPOCC{{\small AFLOW-POCC}}
\def\APL{{\small APL}}
\def\QHAAPL{{\small QHA-APL}}
\def\AAPL{{\small AAPL}}
\def\AFLOWorg{{\small {\sf AFLOW.org}}}
\def\AFLOWSYM{{\small AFLOW-SYM}}
\def\JSON{{\small JSON}}
\def\AFLUX{{\small AFLUX}}
\def\LUX{{\small LUX}}
\def\AURL{{\small AURL}}
\def\AUID{{\small AUID}}

\def\LDA{{\small LDA}}
\def\GGA{{\small GGA}}
\def\PBE{{\small PBE}}
\def\DFT{{\small DFT}}
\def\TDDFT{{\small TDDFT}}
\def\SQS{{\small SQS}}
\def\IFC{{\small IFC}}
\def\API{{\small API}}
\def\RESTAPI{{\small REST-API}}
\def\KPPRA{{\small KPPRA}}
\def\URL{{\small URL}}
\def\GFA{{\small GFA}}
\def\sDebye{{\substack{\scalebox{0.6}{D}}}}

\def\sV{{\substack{\scalebox{0.6}{V}}}}
\def\sDFT{{\substack{\scalebox{0.6}{DFT}}}}

\def\sVRH{{\substack{\scalebox{0.6}{VRH}}}}
\def\sVoigt{{\substack{\scalebox{0.6}{Voigt}}}}
\def\sReuss{{\substack{\scalebox{0.6}{Reuss}}}}

\def\svib{{\substack{\scalebox{0.6}{vib}}}}
\def\sconfig{{\substack{\scalebox{0.6}{config}}}}
\def\sopt{{\substack{\scalebox{0.6}{opt}}}}
\def\sL{{\substack{\scalebox{0.6}{L}}}}
\def\sT{{\substack{\scalebox{0.6}{T}}}}
\def\sB{{\substack{\scalebox{0.6}{B}}}}
\def\sS{{\substack{\scalebox{0.6}{S}}}}

\def\sXC{{\substack{\scalebox{0.6}{XC}}}}

\def\QE{{{\small \textsc{Quantum}ESPRESSO}}}
\def\VASP{{\small VASP}}
\def\ABINIT{{\small ABINIT}}
\def\FHIAIMS{{\small FHI--AIMS}}
\def\SIESTA{{\small SIESTA}}
\def\GAUSSIAN{{\small GAUSSIAN}}
\def\AFLOWML{{\small AFLOW-ML}}

\makeatletter \renewcommand\frontmatter@abstractwidth{\dimexpr\linewidth\relax} \makeatother

\setstcolor{red}

\begin{document}
\title{\Large Automated computation of materials properties}

\author{Cormac Toher}
\affiliation{Department of Mechanical Engineering and Materials Science, Duke University, Durham, North Carolina 27708, United States}
\author{Corey Oses}
\affiliation{Department of Mechanical Engineering and Materials Science, Duke University, Durham, North Carolina 27708, United States}
\author{Stefano Curtarolo}
\email[]{stefano@duke.edu}
\affiliation{Materials Science, Electrical Engineering, Physics and Chemistry, Duke University, Durham North Carolina, 27708, United States}

\date{\today}

\begin{abstract}
\noindent
Materials informatics offers a promising pathway towards rational materials design,
replacing the current trial-and-error approach and accelerating the development of new functional materials.
Through the use of sophisticated data analysis techniques, underlying property trends can be identified,
facilitating the formulation of new design rules.
Such methods require large sets of consistently generated, programmatically accessible materials data.
Computational materials design frameworks using standardized parameter sets
are the ideal tools for producing such data.
This work reviews the state-of-the-art in computational materials design,
with a focus on these automated \textit{ab-initio} frameworks.
Features such as structural prototyping and automated error
correction that enable rapid generation of large datasets are discussed,
and the way in which integrated workflows can simplify the calculation of complex properties,
such as thermal conductivity and mechanical stability, is demonstrated.
The organization of large datasets composed of \textit{ab-initio} calculations, and the tools that render them
programmatically accessible for use in statistical learning applications, are also described.
Finally, recent advances in leveraging existing data to predict novel functional materials,
such as entropy stabilized ceramics, bulk metallic glasses, thermoelectrics, superalloys, and magnets, are surveyed.
\end{abstract}

\maketitle

\section{Introduction}

Materials informatics requires large repositories of materials data to identify trends in and correlations between materials properties,
as well as for training machine learning models.
Such patterns lead to the formulation of descriptors that guide rational materials design.
Generating large databases of computational materials properties requires robust, integrated, automated frameworks \cite{nmatHT}.
Built-in error correction and standardized parameter sets enable the production and analysis of data without direct intervention from human researchers.
Current examples of such frameworks include
\AFLOW\ (\underline{A}utomatic \underline{{\small FLOW}}) \cite{curtarolo:art65, curtarolo:art58, aflowlibPAPER, curtarolo:art92, curtarolo:art104, aflowlib.org, curtarolo:art142, curtarolo:art127, paoflow},
Materials Project \cite{materialsproject.org, APL_Mater_Jain2013, CMS_Ong2012b, Mathew_Atomate_CMS_2017},
OQMD (\underline{O}pen \underline{Q}uantum \underline{M}aterials \underline{D}atabase) \cite{Saal_JOM_2013, Kirklin_AdEM_2013, Kirklin_ActaMat_2016},
the Computational Materials Repository \cite{cmr_repository} and its associated scripting interface ASE (\underline{A}tomic \underline{S}imulation \underline{E}nvironment) \cite{ase},
AiiDA (\underline{A}utomated \underline{I}nteractive \underline{I}nfrastructure and \underline{Da}tabase for Computational Science) \cite{aiida.net, Pizzi_AiiDA_2016, Mounet_AiiDA2D_NNano_2018},
and the Open Materials Database at \verb|httk.openmaterialsdb.se| with its associated \underline{H}igh-\underline{T}hroughput \underline{T}ool\underline{k}it (HTTK).
Other computational materials science resources include the aggregated repository maintained by the \underline{No}vel \underline{Ma}terials \underline{D}iscovery (NoMaD) Laboratory \cite{nomad},
the Materials Mine database available at \verb|www.materials-mine.com|,
and the \underline{T}heoretical \underline{C}rystallography \underline{O}pen \underline{D}atabase (TCOD) \cite{Merkys_TCOD_2017}.
For this data to be consumable by automated machine learning algorithms,
it must be organized in programmatically accessible repositories \cite{aflowlibPAPER, curtarolo:art92, aflowlib.org, materialsproject.org, APL_Mater_Jain2013, Saal_JOM_2013, nomad}.
These frameworks also contain modules that combine and analyze data from various calculations to predict complex thermomechanical phenomena, such as lattice thermal conductivity and mechanical stability.

Computational strategies have already had success in predicting materials for
applications including photovoltaics \cite{YuZunger2012_PRL},
water-splitters \cite{CastelliJacobsen2012_EnEnvSci},
carbon capture and gas storage \cite{LinSmit2012_NMAT_carbon_capture, Alapati_JPCC_2012},
nuclear detection and scintillators \cite{Derenzo:2011io, Ortiz09, aflowSCINT, curtarolo:art46},
topological insulators \cite{curtarolo:art77, Lin_NatMat_HalfHeuslers_2010},
piezoelectrics \cite{Armiento_PRB_2011, Vanderbilt_Piezoelectrics_PRL2012},
thermoelectric materials \cite{curtarolo:art68, madsen2006, curtarolo:art84, curtarolo:art85},
catalysis \cite{Norskov09},
and battery cathode materials \cite{Hautier-JMC2011, Hautier-ChemMater2011, Mueller-ChemMater2011}.
More recently, computational materials data has been combined with machine learning approaches
to predict electronic and thermomechanical properties \cite{curtarolo:art124, deJong_SR_2016},
and to identify superconducting materials \cite{curtarolo:art94}.
Descriptors are also being constructed to describe the formation of disordered
materials, and have recently been used to predict the glass forming
ability of binary alloy systems \cite{curtarolo:art112}.
These successes demonstrate that
accelerated materials design can be achieved by combining structured data sets generated
using autonomous computational methods with intelligently formulated descriptors and machine learning.

\section{Automated computational materials design frameworks}
\label{htframeworks}

\begin{figure*}[t!]
\centering
\includegraphics[width=1.00\linewidth]{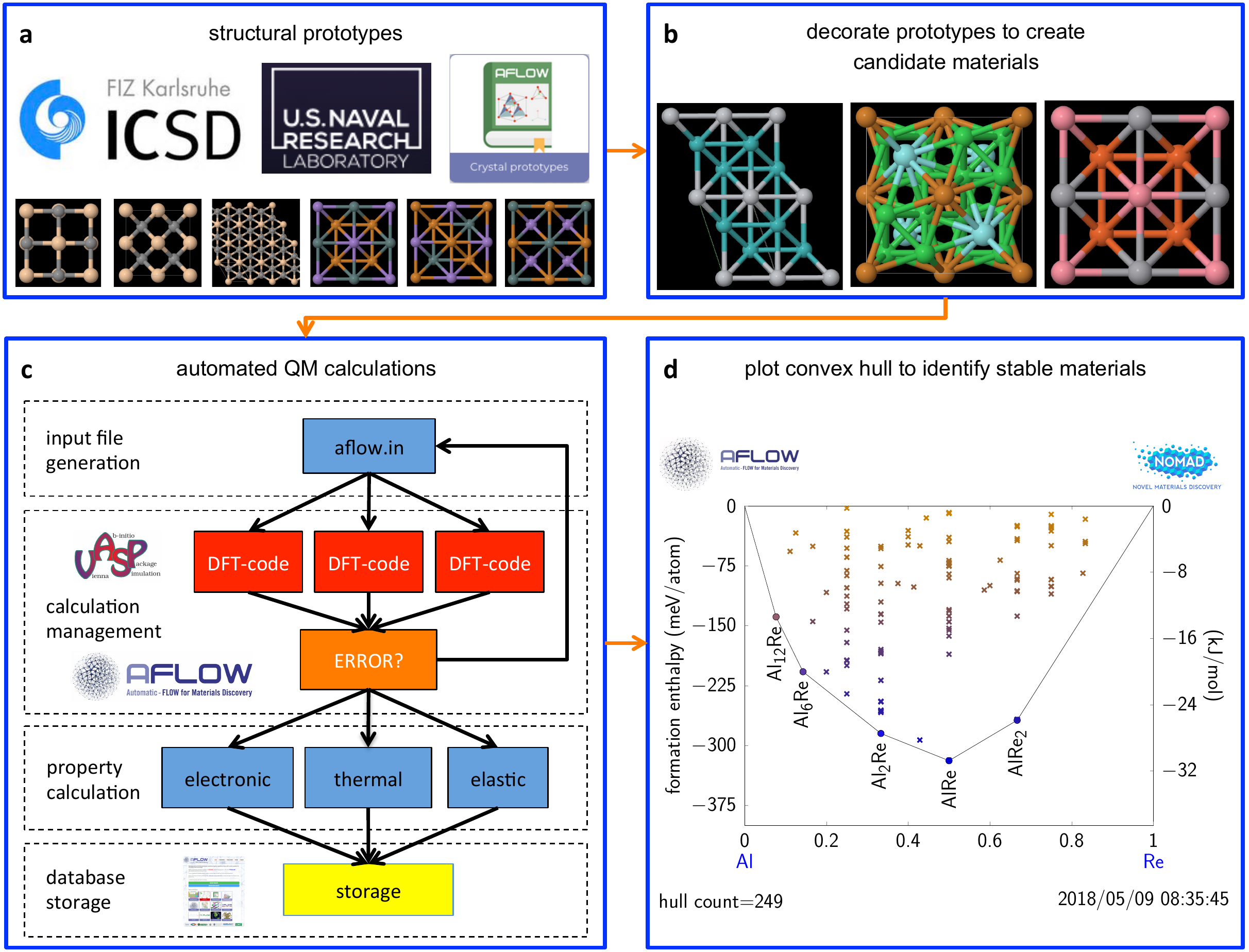}
\vspace{-4mm}
\caption{
\small
Computational materials data generation workflow.
{\bf (a)} Crystallographic prototypes are extracted from databases such as the
ICSD or the NRL crystal structure library, or generated by enumeration algorithms.
The illustrated examples are for the rocksalt,  zincblende, wurtzite, Heusler, anti-Heusler and
half-Heusler structures.
{\bf (b)} New candidate materials are generated by decorating the atomic sites with different elements.
{\bf (c)} Automated \DFT\ calculations are used to optimize the geometric structure and calculate energetic, electronic,
thermal, and elastic properties.
Calculations are monitored to detect errors.
The input parameters are adjusted to compensate for the problem and the calculation is re-run.
Results are formatted and added to an online data repository to facilitate programmatic access.
{\bf (d)} Calculated data is used to plot the convex hull phase diagrams for each alloy system to identify stable compounds.
}
\label{fig:materials_design_workflow}
\end{figure*}

Rapid generation of materials data relies on automated frameworks such as
\AFLOW\ \cite{curtarolo:art65, curtarolo:art58, aflowlibPAPER, curtarolo:art92, curtarolo:art104},
Materials Project's \verb|pymatgen| \cite{CMS_Ong2012b} and \verb|atomate| \cite{Mathew_Atomate_CMS_2017},
OQMD \cite{Saal_JOM_2013, Kirklin_AdEM_2013, Kirklin_ActaMat_2016},
ASE \cite{ase},  and AiiDA \cite{Pizzi_AiiDA_2016}.
The general automated workflow is illustrated in Figure \ref{fig:materials_design_workflow}.
These frameworks begin by creating the input files required by the electronic structure
codes that perform the quantum-mechanics level calculations, where the initial geometry is
generated by decorating structural prototypes (Figure \ref{fig:materials_design_workflow}(a, b)).
They execute and monitor these calculations, reading any error messages written to the
output files and diagnosing calculation failures.
Depending on the nature of the errors, these frameworks are equipped with a catalog of prescribed solutions ---
enabling them to adjust the appropriate parameters and restart the calculations (Figure \ref{fig:materials_design_workflow}(c)).
At the end of a successful calculation, the frameworks parse the output files to extract the relevant materials data
such as total energy, electronic band gap, and relaxed cell volume.
Finally, the calculated properties are organized and formatted for entry into machine-accessible, searchable and sortable databases.

In addition to running and managing the quantum-mechanics level calculations, the frameworks also
maintain a broad selection of post-processing libraries for extracting additional properties,
such as calculating x-ray diffraction (XRD) spectra from relaxed atomic coordinates, and the
formation enthalpies for the convex hull analysis to identify stable compounds (Figure \ref{fig:materials_design_workflow}(d)).
Results from calculations of distorted structures can be combined to calculate
thermal and elastic properties \cite{curtarolo:art65, curtarolo:art96, curtarolo:art100, curtarolo:art115},
and results from different compositions and structural phases can be amalgamated to generate thermodynamic phase diagrams.

\subsection{Generating and using databases for materials discovery}

A major aim of high-throughput computational materials science is to identify new, thermodynamically stable compounds.
This requires the generation of new materials structures, which have not been previously reported in the literature,
to populate the databases. The accuracy of analyses involving sets of structures, such as that used to determine thermodynamic stability,
is contingent on sufficient exploration of the full range of possibilities. Therefore, autonomous materials design frameworks
such as \AFLOW\ use crystallographic prototypes to generate new materials entries consistently and reproducibly.

Crystallographic prototypes are the basic building blocks used to generate the wide range of materials entries involved in
computational materials discovery.
These prototypes are based on \textbf{i.} structures commonly observed in nature \cite{ICSD, navy_crystal_prototypes, curtarolo:art121},
such as the rocksalt, zincblende, wurtzite or Heusler structures illustrated in Figure \ref{fig:materials_design_workflow}(b),
as well as \textbf{ii.} hypothetical structures, such as those enumerated by the methods described in Refs. \onlinecite{enum1, enum2}.
The \AFLOW\ Library of Crystallographic Prototypes \cite{curtarolo:art121} is also available online at \url{aflow.org/CrystalDatabase/}, where
users can choose from hundreds of crystal prototypes with adjustable parameters, and which can be decorated to generate new input
structures for materials science calculations.

New materials are then generated by decorating the various atomic sites in the crystallographic prototype with different elements.
These decorated prototypes serve as the structural input for \textit{ab-initio} calculations.
A full relaxation of the geometries and energy determination follows, from which phase diagrams for stability analyses can be constructed.
The resulting materials data are then stored in an online data repository for future consideration.

\begin{figure}[t!]
\includegraphics[width=1.00\linewidth]{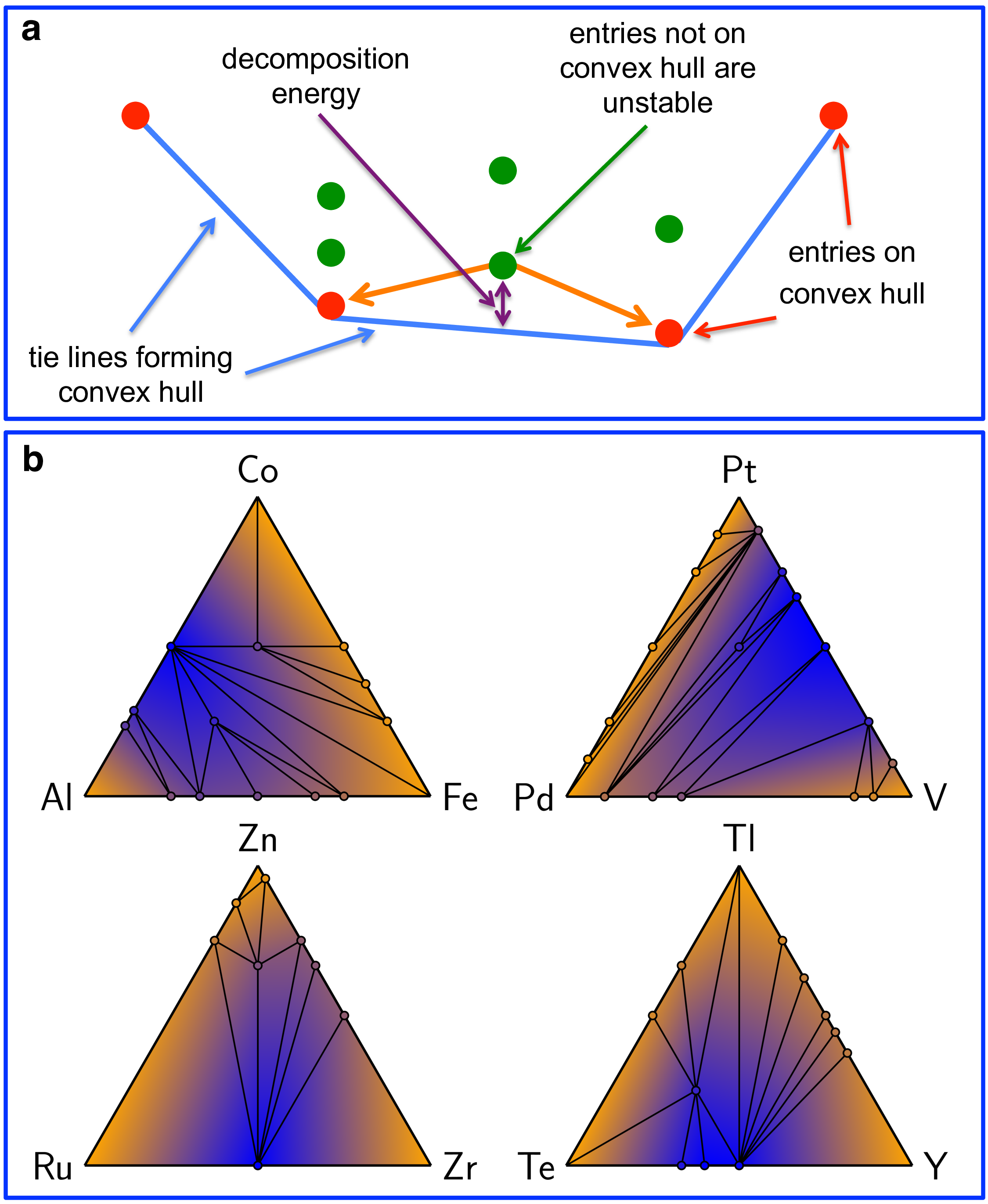}
\vspace{-4mm}
\caption{
\small
Convex hull phase diagrams for multicomponent alloys systems.
{\bf (a)} Schematic illustrating construction of convex hull for a general
  binary alloy system $A_{x}B_{1-x}$. Ground state structures are depicted as red points, with the minimum energy
surface outlined with blue lines. The minimum energy surface is formed by
connecting the lowest energy structures with tie lines which form a convex hull.
Unstable structures are shown in green, with the decomposition reaction indicated
by orange arrows, and the decomposition energy indicated in purple.
{\bf (b)} Example ternary convex hulls as generated by \AFLOW.
}
\label{fig:convex_hulls}
\end{figure}

The phase diagram of a given alloy system can be approximated by considering the low-temperature limit in
which the behavior of the system is dictated by the ground state~\cite{monster, monsterPGM}.
In compositional space, the lower-half convex hull defines the minimum energy surface and the
ground-state configurations of the system.
All non-ground-state stoichiometries are unstable, with the decomposition described by the
hull facet directly below it.
In the case of a binary system, the facet is a tie-line as illustrated in Figure \ref{fig:convex_hulls}(a).
The energy gained from this decomposition is geometrically represented by the (vertical-)distance of the
compound from the facet and quantifies the excitation energy involved in forming this compound.
While the minimum energy surface changes at finite temperature (favoring disordered structures),
the $T=0$~K excitation energy serves as a reasonable descriptor for relative thermodynamic
stability \cite{curtarolo:art113}.
This analysis generates valuable information such as ground-state structures,
excitation energies, and phase coexistence for storage in the
online data repository.
This stability data can be visualized and displayed by online modules,
such as those developed by \AFLOW\ \cite{curtarolo:art113}, the Materials Project \cite{Ong_ChemMat_2008},
and the OQMD~\cite{Akbarzadeh2007, Kirklin_AdEM_2013}.
An example visualization from \AFLOW\ is shown in Figure \ref{fig:convex_hulls}(b).

Convex hull phase diagrams have been used to discover new thermodynamically
stable compounds in a wide range of alloy systems, including hafnium \cite{curtarolo:art49, curtarolo:art51},
rhodium \cite{curtarolo:art53}, rhenium \cite{curtarolo:art63}, ruthenium \cite{curtarolo:art67}, and technetium  \cite{curtarolo:art70}
with various transition metals, as well as the Co-Pt system \cite{curtarolo:art66}. Magnesium alloy systems such as the lightweight
Li-Mg system \cite{curtarolo:art55} and 34 other Mg-based systems \cite{curtarolo:art54} have also been investigated.
This approach has also been used to calculate the solubility of elements in titanium alloys \cite{curtarolo:art47}, to study the effect of hydrogen
on phase separation in iron-vanadium \cite{curtarolo:art74}, and to find new superhard tungsten nitride compounds \cite{curtarolo:art90}.
The data has been employed to generate structure maps for hcp metals \cite{curtarolo:art57},
as well as to search for new stable compounds with the Pt$_8$Ti phase \cite{curtarolo:art56},
and with the $L1_1$ and $L1_3$ crystal structures \cite{curtarolo:art71}.
Note that even if a structure does not lie on the ground state convex hull, this does not rule out its existence.
It may be synthesizable under specific temperature and pressure conditions, and then be metastable under ambient
conditions.

\subsection{Standardized protocols for automated data generation}

Standard calculation protocols and parameters sets \cite{curtarolo:art104} are essential to
the identification of trends and correlations among materials properties.
The workhorse method for calculating quantum-mechanically resolved materials properties
is \underline{d}ensity \underline{f}unctional \underline{t}heory (\DFT).
\DFT\ is based on the Hohenberg-Kohn theorem \cite{Hohenberg_PR_1964}, which proves that for a ground state system,
the potential energy is a unique functional of the density: $V (\vec{r}) = V(\rho(\vec{r}))$.
This allows for the charge density $\rho(\vec{r})$ to be used as the central variable for the calculations
rather than the many-body wave function $\Psi(\vec{r_1}, \vec{r_2}, ..., \vec{r_N})$,
dramatically reducing the number of degrees of freedom in the calculation.

The Kohn-Sham equations \cite{DFT} map the $n$ coupled equations for the system of $n$ interacting particles
onto a system of $n$ independent equations for $n$ non-interacting particles:
\begin{equation}
\label{kohnshameqns}
\left[ -\frac{\hbar^2}{2m} \nabla^2 + V_s (\vec{r}) \right] \phi_i (\vec{r}) = \varepsilon_i \phi_i(\vec{r}),
\end{equation}
where $\phi_i(\vec{r})$ are the non-interacting Kohn-Sham eigenfunctions and $\varepsilon_i$ are their eigenenergies.
$V_s (\vec{r})$ is the Kohn-Sham potential:
\begin{equation}
\label{kohnshampotential}
V_s (\vec{r}) = V(\vec{r}) + \int e^2 \frac{\rho_s (\vec{r'})}{|\vec{r} - \vec{r'}|}
d^3 \vec{r'} + V_\sXC\left[\rho_s(\vec{r})\right],
\end{equation}
where $V(\vec{r})$ is the external potential
(which includes influences of the nuclei, applied fields, and the core electrons when pseudopotentials are used),
the second term is the direct Coulomb potential, and $V_\sXC\left[\rho_s(\vec{r})\right]$ is the exchange-correlation term.

The mapping onto a system of $n$ non-interacting particles comes at the cost of introducing the
exchange-correlation potential $V_\sXC\left[\rho_s(\vec{r})\right]$, the exact form of which is unknown and must be approximated.
The simplest approximation is the \underline{l}ocal \underline{d}ensity \underline{a}pproximation (\LDA) \cite{PerdewZunger},
in which the magnitude of the exchange-correlation energy at a particular point in space is
assumed to be proportional to the magnitude of the density at that point in space.
Despite its simplicity, \LDA\ produces realistic results for atomic structure, elastic and vibrational properties
for a wide range of systems. However, it tends to overestimate the binding energies of materials, even
putting crystal bulk phases in the wrong energetic order \cite{Zupan_LDAperformance_PRB_1998}.
Beyond \LDA\ is the \underline{G}eneralized \underline{G}radient \underline{A}pproximation (\GGA), in which the exchange correlation term
is a functional of the charge density and its gradient at each point in space.
There are several forms of \GGA\, including those developed by Perdew, Burke and Ernzerhof (\PBE, \cite{PBE}), or by Lee, Yang and Parr ({\small LYP}, \cite{LYP_1988}).
A more recent development is the meta-\GGA\ \underline{S}trongly \underline{C}onstrained and \underline{A}ppropriately \underline{N}ormed ({\small SCAN})
functional \cite{Perdew_SCAN_PRL_2015}, which satisfies all 17 known exact constraints on meta-\GGA\ functionals.

The major limitations of \LDA\ and \GGA\ include their inability to adequately describe systems with strongly correlated or localized electrons,
due to the local and semilocal nature of the functionals.
Treatments include the Hubbard $U$ corrections \cite{LiechDFTU, DudarevDFTU}, self-interaction corrections \cite{PerdewZunger}
and hybrid functionals such as Becke's 3-parameter modification of {\small LYP} ({\small B3LYP}, \cite{B3LYP_1993}), and that of Heyd, Scuseria and Ernzerhof ({\small HSE}, \cite{HSE}).

Within the context of \textit{ab-initio} structure prediction calculations, \GGA-\PBE\ is the usual standard since it tends to produce
accurate geometries and lattice constants \cite{monster}.
For accounting for strong correlation effects, the \DFT$+U$ method \cite{LiechDFTU, DudarevDFTU}
is often favored in large-scale automated database generation due to its low computational overhead.
However, the traditional \DFT$+U$ procedure requires the addition of an empirical factor to the potential \cite{LiechDFTU, DudarevDFTU}.
Recently, methods have been implemented to calculate the $U$ parameter self-consistently from first-principles, such as the ACBN0 functional \cite{curtarolo:art93}.

\DFT\ also suffers from an inadequate description of excited/unoccupied states, as the theory
is fundamentally based on the ground state.
Extensions for describing excited states include time-dependent \DFT\ (\TDDFT) \cite{Hedin_GW_1965} and the GW correction \cite{GW}.
However, these methods are typically much more expensive than standard \DFT, and are not generally considered for large scale database generation.

At the technical implementation level, there are
many \DFT\ software packages available, including
\VASP\ \cite{vasp_prb1993, vasp_prb1996, kresse_vasp_1, kresse_vasp_paw},
\QE\ \cite{qe, Giannozzi:2017io}, \ABINIT\ \cite{gonze:abinit, abinit_2009},
\FHIAIMS\ \cite{Blum_CPC2009_AIM}, \SIESTA\ \cite{Soler2002SIESTA} and \GAUSSIAN\ \cite{Gaussian_2009}.
These codes are generally distinguished by the choice of basis set.
There are two principle types of basis sets: plane waves, which take the form $\psi (\vec{r}) = \sum e^{i \vec{k}\cdot\vec{r}}$,
and local orbitals, formed by a sum over functions $\phi_a (\vec{r})$ localized at particular points in space, such as
gaussians or numerical atomic orbitals \cite{Hehre_self_consistent_molecular_orbit_JCP1969}.
Plane wave based packages include \VASP, \QE\ and \ABINIT, and are generally better suited to periodic systems such as bulk inorganic materials.
Local orbital based packages include \FHIAIMS, \SIESTA\ and \GAUSSIAN, and are generally better suited to non-periodic systems such as organic molecules.
In the field of automated computational materials science, plane wave codes such as \VASP\ are generally preferred:
it is straightforward to automatically and systematically generate well-converged basis sets
since there is only a single parameter to adjust, namely the cut-off energy determining the number
of plane waves in the basis set.
Local orbital basis sets tend to have far more independently adjustable degrees of freedom,
such as the number of basis orbitals per atomic orbital as well as their respective cut-off radii,
making the automated generation of reliable basis sets more difficult.
Therefore, a typical standardized protocol for automated materials science calculations \cite{curtarolo:art104} relies on
the \VASP\ software package with a basis set cut-off energy higher than that recommended by the \VASP\ potential files,
in combination with the \PBE\ formulation of \GGA.

\begin{figure*}[t!]
\includegraphics[width=1.00\linewidth]{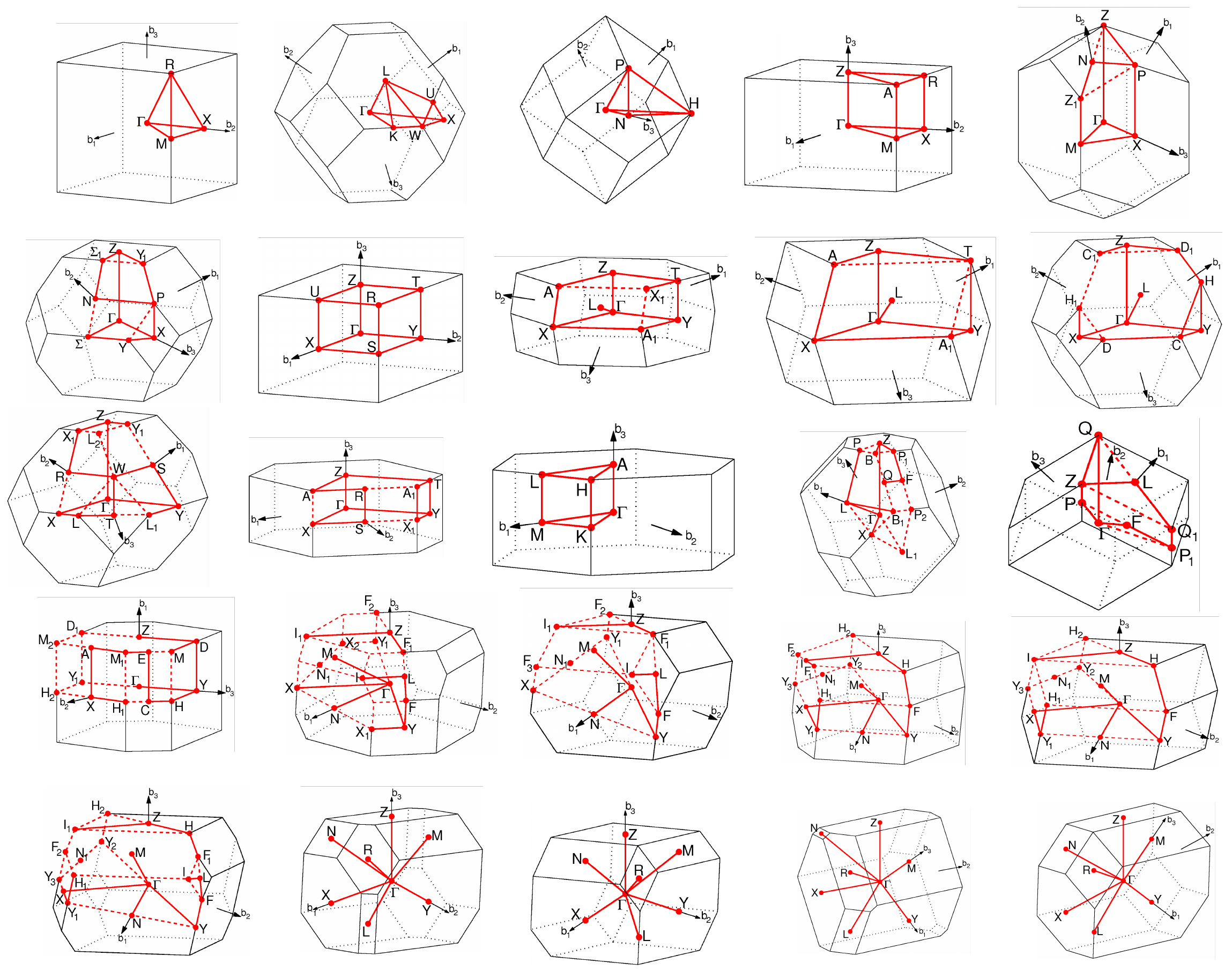}
\vspace{-4mm}
\caption{
\small
Standardized paths in reciprocal space for calculation of the electronic band
structures for the 25 different lattice types \cite{curtarolo:art58}.
}
\label{fig:band_structure_paths}
\end{figure*}

Finally, it is necessary to automate the generation of the $k$-point grid and pathways in reciprocal space used for the calculation of forces, energies and the electronic band structure.
In general, \DFT\ codes use standardized methods such as the Monkhorst-Pack scheme \cite{MonkhorstPack} to generate reciprocal lattice $k$-point grids,
although optimized grids have been calculated for different lattice types and are available online \cite{Wisesa_Kgrids_PRB_2016}.
Optimizing $k$-point grid density is a computationally expensive process that is difficult to automate,
so instead standardized grid densities based on the concept of ``\underline{$k$}-\underline{p}oints \underline{p}er \underline{r}eciprocal \underline{a}tom'' (\KPPRA) are used.
The \KPPRA\ value is chosen to be sufficiently large to ensure convergence for all systems.
Typical recommended values used for \KPPRA\ range from 6,000 to 10,000 \cite{curtarolo:art104},
so that a material with two atoms in the calculation cell will have a $k$-point mesh of at least 3,000 to 5,000 points.
Standardized directions in reciprocal space have also been defined for the calculation of the band structure as illustrated in Figure \ref{fig:band_structure_paths} \cite{curtarolo:art58}.
These paths are optimized to include all of the high-symmetry points of the lattice.

\section{Integrated calculation of materials properties}
\label{sect:thermomechanical}

Automated frameworks such as \AFLOW\ combine the computational analysis of properties including symmetry, electronic structure,
elasticity, and thermal behavior into integrated workflows.
Crystal symmetry information is used to find the primitive cell to reduce the size of \DFT\ calculations,
to determine the appropriate paths in reciprocal space for electronic band structure calculations (see Figure \ref{fig:band_structure_paths}, \cite{curtarolo:art58}),
and to determine the set of inequivalent distortions for phonon and elasticity calculations.
Thermal and elastic properties of materials are important for predicting the thermodynamic and mechanical stability
of structural phases \cite{Greaves_Poisson_NMat_2011, Poirier_Earth_Interior_2000, Mouhat_Elastic_PRB_2014, curtarolo:art106}
and assessing their importance for a variety of applications.
Elastic properties such as the shear and bulk moduli are important for predicting the hardness
of materials \cite{Chen_hardness_Intermetallics_2011, Teter_Hardness_MRS_1998},
and thus their resistance to wear and distortion.
Elasticity tensors can be used to predict the properties of composite
materials \cite{Hashin_Multiphase_JMPS_1963, Zohdi_Polycrystalline_IJNME_2001}.
They are also important in geophysics for modeling the propagation of seismic waves
in order to investigate the mineral composition of geological
formations \cite{Poirier_Earth_Interior_2000, Anderson_Elastic_RGP_1968, Karki_Elastic_RGP_2001}.
The lattice thermal conductivity $\left(\kappa_\sL\right)$ is a crucial
design parameter in a wide range of important
technologies, such as the development of new thermoelectric
materials \cite{zebarjadi_perspectives_2012,curtarolo:art84,Garrity_thermoelectrics_PRB_2016},
heat sink materials for thermal management in electronic devices \cite{Yeh_2002},
and rewritable phase-change memories \cite{Wright_tnano_2011}.
High thermal conductivity materials, which typically have a zincblende or diamond-like structure, are essential
in microelectronic and nanoelectronic devices for achieving
efficient heat removal \cite{Watari_MRS_2001}, and have
been intensively studied for the past few decades \cite{Slack_1987}.
Low thermal conductivity materials constitute
the basis of a new generation of thermoelectric materials and thermal
barrier coatings \cite{Snyder_jmatchem_2011}.

The calculation of thermal and elastic properties offer an excellent example of the power of
integrated computational materials design frameworks.
With a single input file, these frameworks can automatically set-up and run calculations of
different distorted cells, and combine the resulting energies and
forces to calculate thermal and mechanical properties.

\begin{figure*}
  \includegraphics[width=1.00\linewidth]{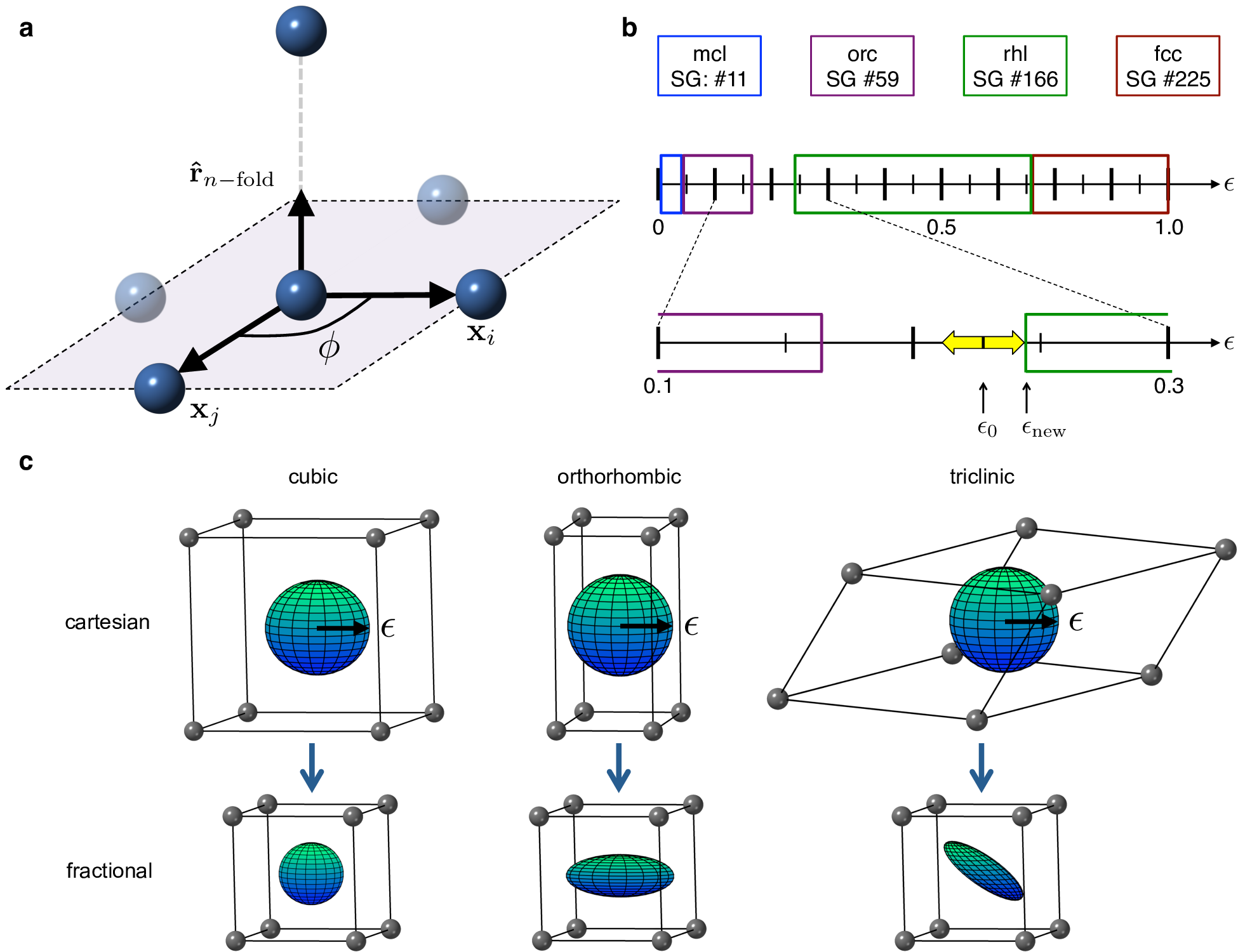}
  \vspace{-4mm}
  \caption{\small \textbf{Challenges in autonomous symmetry analysis.}
    (\textbf{a}) An illustration of a general $n$-fold symmetry operation.
    (\textbf{b}) Possible space group determinations with mapping tolerance $\epsilon$ for AgBr (ICSD \#56551).
    (\textbf{c}) Warping of mapping tolerance sphere with a transformation from cartesian to fractional basis.
    }
  \label{fig:sym}
\end{figure*}

\subsection{Autonomous symmetry analysis}

Critical to any analysis of crystals is the accurate determination of the symmetry profile.
For example, symmetry serves to
\textbf{i.} validate the forms of the elastic constants
and compliance tensors, where the crystal symmetry dictates equivalence or absence
of specific tensor elements~\cite{nye_symmetry, curtarolo:art100, Mouhat_Elastic_PRB_2014}, and
\textbf{ii.} reduce the number of \textit{ab-initio} calculations needed for phonon
calculations, where, in the case of the finite-displacement method, equivalent
atoms and distortion directions are identified through factor group and site symmetry
analyses~\cite{Maradudin1971}.

Autonomous workflows for elasticity and vibrational characterizations
therefore require a correspondingly robust symmetry analysis.
Unfortunately, standard symmetry packages~\cite{stokes_findsym,Stokes_FROZSL_Ferroelectrics_1995,platon_2003,spglib},
catering to different objectives, depend on tolerance-tuning to
overcome numerical instabilities and atypical data --- emanating from
finite temperature measurements and uncertainty in experimentally reported observations.
These tolerances are responsible for validating mappings and identifying isometries,
such as the $n$-fold operator depicted in Figure~\ref{fig:sym}(a).
Some standard packages define separate tolerances for space, angle~\cite{spglib},
and even operation type~\cite{stokes_findsym,Stokes_FROZSL_Ferroelectrics_1995,platon_2003}
(\textit{e.g.}, rotation \textit{versus} inversion).
Each parameter introduces a factorial expansion of unique inputs, which can result in
distinct symmetry profiles as illustrated in Figure~\ref{fig:sym}(b).
By varying the spatial tolerance $\epsilon$, four different space groups can be observed
for AgBr (ICSD \#56551 \footnote{{h}ttp://www.aflow.org/material.php?id=56551}), if one is found at all.
Gaps in the range, where no consistent symmetry profile can be resolved, are
particularly problematic in automated frameworks, triggering critical failures in subsequent analyses.

Cell shape can also complicate mapping determinations.
Anisotropies in the cell, such as skewness of lattice vectors, translate
to distortions of fractional and reciprocal spaces.
A uniform tolerance sphere in cartesian space, inside which points are considered mapped,
generally warps to a sheared spheroid, as depicted in Figure~\ref{fig:sym}(c).
Hence, distances in these spaces are direction-dependent, compromising the integrity
of rapid minimum-image determinations~\cite{hloucha_minimumimage_1998} and generally warranting
prohibitively expensive algorithms~\cite{curtarolo:art135}.
Such failures can result in incommensurate symmetry profiles, where the real space
lattice profile (\textit{e.g.}, bcc) does not match that of the reciprocal space (fcc).

The new \AFLOWSYM\ module \cite{curtarolo:art135} within \AFLOW\ offers careful treatment of tolerances, with extensive
validation schemes, to mitigate the aforementioned challenges.
Although a user-defined tolerance input is still available, \AFLOW\ defaults to one of two pre-defined
tolerances, namely \texttt{tight} (standard) and \texttt{loose}.
Should any discrepancies occur, these defaults are the starting values of a large tolerance scan,
as shown in Figure~\ref{fig:sym}(b).
A number of validation schemes have been incorporated to catch such discrepancies.
These checks are consistent with crystallographic group theory principles, validating operation
types and cardinalities~\cite{tables_crystallography}.
From considerations of different extreme cell shapes, a heuristic threshold has been defined
to classify scenarios where mapping failures are likely to occur --- based on skewness and mapping tolerance.
When benchmarked against standard packages for over 54,000 structures in the Inorganic Crystal Structure Database,
\AFLOWSYM\ consistently resolves
the symmetry characterization most compatible with experimental observations~\cite{curtarolo:art135}.

Along with accuracy, \AFLOWSYM\ delivers a wealth of symmetry properties and representations
to satisfy injection into any analysis or workflow.
The full set of operators --- including that of the point-, factor-, crystallographic point-, space groups,
and site symmetries --- are provided in matrix, axis-angle, matrix generator, and quaternion representations in
both cartesian and fractional coordinates.
A span of characterizations, organized by degree of symmetry-breaking, are available, including
those of the lattice, superlattice, crystal, and crystal-spin.
Space group and Wyckoff positions are also resolved.
The full dataset is made available in both plain-text and \JSON\ formats.

\begin{figure*}[t!]
\centering
\includegraphics[width=1.00\linewidth]{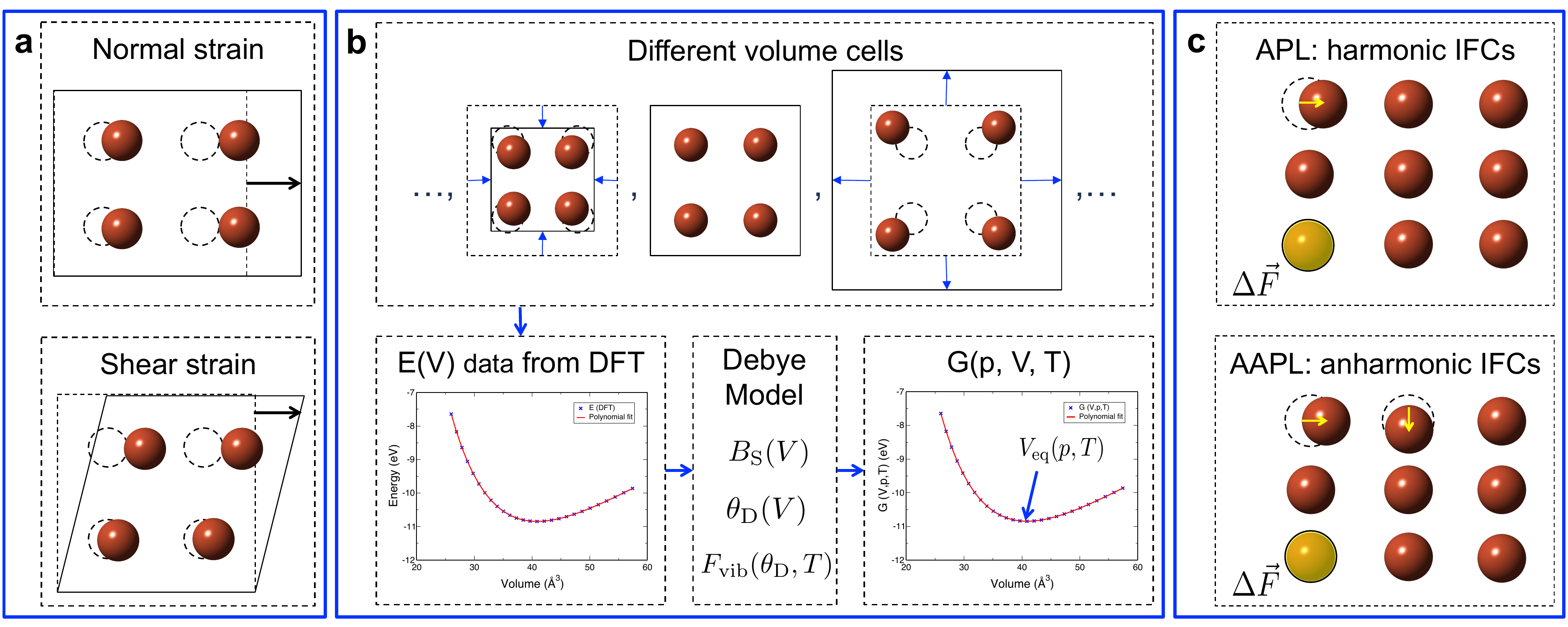}
\vspace{-4mm}
\caption{
\small
{\bf (a)} \AEL\ applies a set of independent normal and shear strains to the crystal structure to obtain
the elastic constants.
{\bf (b)} \AGL\ applies a set of isotropic strains to the unit cell to obtain
energy \textit{versus} volume data, which is fitted by a polynomial in order to
calculate the bulk modulus as a function of volume, $B_{\mathrm S} (V)$.
$B_{\mathrm S} (V)$ is then used to calculate the Debye temperature as a function of volume and thus
the vibrational free energy as a function of temperature.
The Gibbs free energy as a function of volume is then minimized for each pressure and temperature
point to obtain the equilibrium volume and other thermomechanical properties.
{\bf (c)} APL obtains the harmonic \underline{i}nteratomic \underline{f}orce \underline{c}onstants (\IFC{}s) from supercell calculations where inequivalent
atoms are displaced in inequivalent directions, and then the changes in the forces on the other atoms are calculated.
The \IFC{}s are then used to construct the dynamical matrix, which is diagonalized to obtain the phonon eigenmodes.
AAPL calculates three-phonon scattering effects by performing supercell calculations where pairs of inequivalent atoms are displaced in inequivalent directions, and
 the changes in the forces on the other atoms in the supercell are calculated to obtain the third-order anharmonic \IFC{}s.
}
\label{fig:thermomechanical}
\end{figure*}

\subsection{Elastic Constants}

There are two main methods for calculating the elastic constants
based on the response of either the stress tensor or the total energy
to a set of applied strains
\cite{curtarolo:art100, curtarolo:art115, Golesorkhtabar_ElaStic_CPC_2013, Silveira_Elastic_CPC_2008, Silveira_Elastic_CPC_2008, Silva_Elastic_PEPI_2007}.
Automated implementations of these methods are included in the \AFLOW\
(referred to as the \underline{A}utomatic \underline{E}lasticity \underline{L}ibrary, \AEL\ \cite{curtarolo:art115})
and Materials Project frameworks \cite{curtarolo:art100}.

To calculate the elastic tensor
several different normal and shear strains should be applied to the calculation cell in each
independent direction \cite{curtarolo:art100, curtarolo:art115}, as illustrated in Figure \ref{fig:thermomechanical}(a).
The resulting stress tensor elements $\sigma_{ij}$, obtained from the directional forces on the cell calculated with \DFT, can then be fitted to the applied strains $\epsilon_{ij}$
to obtain the corresponding elastic constants $c_{ij}$ in the form of the stiffness tensor:
\begin{equation}
\left( \begin{array}{l} \sigma_{11} \\ \sigma_{22} \\ \sigma_{33} \\ \sigma_{23} \\ \sigma_{13} \\ \sigma_{12} \end{array} \right) =
\left( \begin{array}{l l l l l l} c_{11}\ c_{12}\ c_{13}\ c_{14}\ c_{15}\ c_{16} \\
c_{12}\ c_{22}\ c_{23}\ c_{24}\ c_{25}\ c_{26} \\
c_{13}\ c_{23}\ c_{33}\ c_{34}\ c_{35}\ c_{36} \\
c_{14}\ c_{24}\ c_{34}\ c_{44}\ c_{45}\ c_{46} \\
c_{15}\ c_{25}\ c_{35}\ c_{45}\ c_{55}\ c_{56} \\
c_{16}\ c_{26}\ c_{36}\ c_{46}\ c_{56}\ c_{66} \end{array} \right)
\left( \begin{array}{c} \epsilon_{11} \\ \epsilon_{22} \\ \epsilon_{33} \\ 2\epsilon_{23} \\
2\epsilon_{13} \\ 2\epsilon_{12} \end{array} \right)
\end{equation}
written in the $6\times 6$ Voigt notation using the mapping \cite{Poirier_Earth_Interior_2000}:
$11 \mapsto 1$, $22 \mapsto 2$, $33 \mapsto 3$, $23 \mapsto 4$, $13 \mapsto 5$, $12 \mapsto 6$.
Symmetry analysis such as that provided by \AFLOWSYM\ can be used to reduce the number of required calculations
by up to a factor of three in the case of cubic systems, as well as for verification of the computed tensors \cite{Mouhat_Elastic_PRB_2014}.

The elastic constants can then be used in the Voigt or Reuss approximations,
which for polycrystalline materials correspond to assuming uniform strain and uniform stress respectively,
and give the upper and lower bounds on the elastic moduli.
In the Voigt approximation, the bulk modulus is given by
\begin{equation}
\label{bulkmodvoigt}
B_{\sVoigt} = \frac{1}{9} \left[ (c_{11} + c_{22} + c_{33}) + 2 (c_{12} + c_{23} + c_{13}) \right],
\end{equation}
and the shear modulus is given by
\begin{multline}
\label{shearmodvoigt}
G_{\sVoigt} = \frac{1}{15} \left[ (c_{11} + c_{22} + c_{33}) -  (c_{12} + c_{23} + c_{13}) \right] + \\
\frac{1}{5} (c_{44} + c_{55} + c_{66}).
\end{multline}
The Reuss approximation uses the elements of the compliance tensor $s_{ij}$ (the inverse of the stiffness tensor)
to calculate the bulk modulus
\begin{equation}
\label{bulkmodreuss}
\frac{1}{B_{\sReuss}} =  (s_{11} + s_{22} + s_{33}) + 2 (s_{12} + s_{23} + s_{13}),
\end{equation}
while the shear modulus is given by
\begin{multline}
\label{shearmodreuss}
\frac{15}{G_{\sReuss}} = 4(s_{11} + s_{22} + s_{33}) - 4 (s_{12} + s_{23} + s_{13}) + \\
+ 3 (s_{44} + s_{55} + s_{66}).
\end{multline}
The two approximations are combined to obtain the \underline{V}oigt-\underline{R}euss-\underline{H}ill (\VRH)
averages \cite{Hill_elastic_average_1952} for the bulk modulus
\begin{equation}
\label{bulkmodvrh}
B_{\sVRH} = \frac{B_{\sVoigt} + B_{\sReuss}}{2};
\end{equation}
and the shear modulus
\begin{equation}
\label{shearmodvrh}
G_{\sVRH} = \frac{G_{\sVoigt} + G_{\sReuss}}{2}.
\end{equation}
The Poisson ratio $\nu$ is then given by
\begin{equation}
\label{Poissonratio}
\nu = \frac{3 B_{\sVRH} - 2 G_{\sVRH}}{6 B_{\sVRH} + 2 G_{\sVRH}}.
\end{equation}

\subsection{Quasi-harmonic Debye-Gr{\"u}neisen model}
\label{sect:qha_debye}

Thermal properties can be predicted by several different methods, such as the quasi-harmonic Debye-Gr{\"u}neisen
model which uses volume as a proxy for temperature \cite{curtarolo:art96},
and by calculating the phonon dispersion from the dynamical matrix of \IFC{}s~\cite{Maradudin1971}.

The energy \textit{versus} volume data from a set of simple static primitive cell calculations can be
fitted to a quasi-harmonic Debye-Gr{\"u}neisen model such as the ``GIBBS'' method \cite{Blanco_CPC_GIBBS_2004, curtarolo:art96, curtarolo:art115}
to obtain thermal properties, as demonstrated in Figure \ref{fig:thermomechanical}(b). This method has been implemented in the \AFLOW\ framework
in the form of the \underline{A}utomatic \underline{G}IBBS \underline{L}ibrary (\AGL).

First, the adiabatic bulk modulus $B_\sS$ as a function of cell volume $V$ is obtained either
\textbf{i.} by fitting the $E_\sDFT(V)$ data to an equation of state (\EOS), or
\textbf{ii.} by taking the numerical second
derivative of a polynomial fit of $E_\sDFT(V)$, which gives the static bulk modulus $B_{\mathrm{static}}$:
\begin{eqnarray}
\label{bulkmod}
B_\sS (V) &\approx& B_{\mathrm{static}} (\vec{x}) \approx B_{\mathrm{static}}(\vec{x}_\sopt(V)) =\\ \nonumber
&=&V \left( \frac{\partial^2 E(\vec{x}_\sopt (V))}{\partial V^2} \right) = V \left( \frac{\partial^2 E(V)}{\partial V^2} \right).
\end{eqnarray}
Three different empirical \EOS\ have been implemented within \AGL:
the Birch-Murnaghan \EOS\ \cite{Birch_Elastic_JAP_1938, Poirier_Earth_Interior_2000, Blanco_CPC_GIBBS_2004};
the Vinet \EOS\ \cite{Vinet_EoS_JPCM_1989, Blanco_CPC_GIBBS_2004};
and the Baonza-C{\'a}ceres-N{\'u}{\~n}ez spinodal \EOS\ \cite{Baonza_EoS_PRB_1995, Blanco_CPC_GIBBS_2004}.
However, these \EOS\ often introduce an additional source of error into the
results since they are calibrated
for specific sets of systems and pressure-temperature regimes.
Recent studies have found the numerical calculation of $B$ to be just as, if not more, reliable as the empirical \EOS\ \cite{curtarolo:art115}.
Therefore, the numerical method is the default for the automated generation of thermomechanical properties for the \AFLOW\ database.

The bulk modulus can then be used to calculate the Debye temperature as a function of volume:
\begin{equation}
\label{debyetemp}
\theta_\sDebye (V) = \frac{\hbar}{k_\sB}[6 \pi^2 V^{1/2} n]^{1/3} f(\nu) \sqrt{\frac{B_\sS}{M}},
\end{equation}
where $M$ is the mass of the unit cell, and $f(\nu)$ is a function of the Poisson ratio $\nu$:
\begin{equation}
\label{fpoisson}
f(\nu) = \left\{ 3 \left[ 2 \left( \frac{2}{3} \!\cdot\! \frac{1 + \nu}{1 - 2 \nu} \right)^{3/2}
+ \left( \frac{1}{3} \!\cdot\! \frac{1 + \nu}{1 - \nu} \right)^{3/2} \right]^{-1} \right\}^{\frac{1}{3}}\!\!\!\!.
\end{equation}
The integration offered by the \AFLOW\ framework allows the value of $\nu$ required by this expression to be obtained directly and automatically
from the \AEL\ calculation (Equation \ref{Poissonratio}).

To obtain the equilibrium volume at a particular $(p, T)$ point, the Gibbs free energy is minimized with
respect to volume.
In the quasi-harmonic approximation, the vibrational component of the free energy, $F_\svib(\vec{x}; T)$,
is given by
\begin{equation}
F_\svib(\vec{x}; T) \!=\!\! \int_0^{\infty} \!\!\left[\frac{\hbar \omega}{2} \!+\!
  \frac{1}{\beta} \ \mathrm{log}\!\left(1\!-\!{\mathrm e}^{- \beta \hbar \omega }\right)\!\right]\!g(\vec{x}; \omega) d\omega,
  \label{eq:fvib}
\end{equation}
where $\beta=\left(k_\sB T\right)^{-1}$ and $g(\vec{x}; \omega)$ is the phonon density of states, which depends on the system geometry $\vec{x}$.
In the Debye-Gr{\"u}neisen model, $F_\svib$ can be written as
\begin{equation}
\label{helmholtzdebye}
  F_\svib(\theta_\sDebye; T) \!=\! \frac{n}{\beta} \!\left[ \frac{9}{8} \frac{\theta_\sDebye}{T} \!+\! 3\ \mathrm{log}\!\left(1 \!-\!
{\mathrm e}^{- \theta_\sDebye / T}\!\right) \!\!-\!\! D\left(\frac{\theta_\sDebye}{T}\right)\!\!\right],
\end{equation}
where $D(\theta_\sDebye / T)$ is the Debye integral
\begin{equation}
D \left(\theta_\sDebye/T \right) = 3 \left( \frac{T}{\theta_\sDebye} \right)^3 \int_0^{\theta_\sDebye/T} \frac{x^3}{e^x - 1} dx.
\end{equation}
Next, the full Gibbs free energy as a function of temperature and pressure is calculated
\begin{equation}
\label{gibbsdebye}
{\sf G}(V; p, T) = E_\sDFT(V) + F_{\mathrm{vib}} (\theta_\sDebye(V); T)  + pV,
\end{equation}
and fitted by a polynomial in $V$, the minimum of which gives the equilibrium volume, $V_{\mathrm{eq}}$.
Note that the symbol $G$ is used for shear modulus while {\sf G} is used for the Gibbs free energy.

$\theta_\sDebye$ is then determined from its value at $V_{\mathrm{eq}}$, while
other thermal properties such as the Gr{\"u}neisen parameter can be calculated using the expression
\begin{equation}
\label{gruneisen_theta}
\gamma = - \frac{V}{\theta_\sDebye} \frac{\partial \theta_\sDebye(V)}{\partial V}.
\end{equation}
The specific heat capacity at constant volume can be obtained using the expression
\begin{equation}
\label{heat_capacity_Cv}
C_{\sV, \svib} = 3n k_\sB \left[ 4 D \left(\frac{\theta_\sDebye}{T} \right) - \frac{3 \theta_\sDebye / T}{e^{\theta_\sDebye / T} - 1} \right],
\end{equation}
while the specific heat capacity at constant pressure is given by
\begin{equation}
\label{heat_capacity_Cp}
C_{{\rm p}, \svib} = C_{\sV, \svib} (1 + \alpha \gamma T),
\end{equation}
where $\alpha$ is the coefficient of thermal expansion
\begin{equation}
\label{heat_capacity_Cp_alpha}
\alpha = \frac{\gamma C_{\sV, \svib}}{B_\sT V}.
\end{equation}

The lattice thermal conductivity can be calculated using the
Leibfried-Schl{\"o}mann equation \cite{Leibfried_formula_1954, slack, Morelli_Slack_2006}
using the Debye temperature and the Gr{\"u}neisen parameter
\begin{eqnarray}
\label{thermal_conductivity}
\kappa_\sL (\theta_\acoustic) &=& \frac{0.849 \times 3 \sqrt[3]{4}}{20
\pi^3(1 - 0.514\gamma_\acoustic^{-1} + 0.228\gamma_\acoustic^{-2})}  \\ \nonumber
& &\times \left( \frac{k_\sB \theta_\acoustic}{\hbar} \right)^2
\frac{k_\sB m V^{\frac{1}{3}}}{\hbar \gamma_\acoustic^2}.
\end{eqnarray}
where $V$ is the volume of the unit cell, $m$ is the average atomic mass, while $\theta_\acoustic$ and $\gamma_\acoustic$
are the acoustic Debye temperature and Gr{\"u}neisen parameter obtained by only considering the acoustic modes,
based on the assumption that the optical phonon modes in crystals do not contribute to heat transport
\cite{slack, Morelli_Slack_2006}.
$\theta_\acoustic$ and $\gamma_\acoustic$ can be derived directly from the phonon DOS by only considering the acoustic
modes \cite{slack,Wee_Fornari_TiNiSn_JEM_2012}.
$\theta_\acoustic$ can also be estimated from the traditional Debye temperature $\theta_\sDebye$ using the expression $\theta_\acoustic = \theta_\sDebye n^{-\frac{1}{3}}$ \cite{slack, Morelli_Slack_2006}.
There is no simple way to extract $\gamma_\acoustic$ from the traditional Gr{\"u}neisen parameter, so the approximation
$\gamma_\acoustic = \gamma$ is used in the \AEL-\AGL\ approach to calculating the thermal conductivity.

The thermal conductivity at temperatures other than $\theta_\acoustic$ is estimated
using the expression \cite{slack, Morelli_Slack_2006, Madsen_PRB_2014}:
$\kappa_\sL(T) = \kappa_\sL(\theta_\acoustic) \theta_\acoustic / T$.

\subsection{Harmonic Phonons}

Thermal properties can also be obtained by directly
calculating the phonon dispersion from the dynamical matrix of \IFC{}s.
The approach is implemented within the {\small \underline{A}FLOW} \underline{P}honon \underline{L}ibrary
(\APL)~\cite{curtarolo:art65}.
The \IFC{}s are determined from a set of supercell calculations in which the atoms are
displaced from their equilibrium positions \cite{Maradudin1971} as shown in Figure \ref{fig:thermomechanical}(c).

The \IFC{}s derive from a Taylor expansion of the potential energy, $V$, of the crystal
about the atoms' equilibrium positions:
\begin{multline}
  V=\left.V\right|_{\vec{r}(i,t)=0,\forall i}+
  \sum_{i,\alpha}\left.\frac{\partial V}{\partial r(i,t)^{\alpha}}\right|_{\vec{r}(i,t)=0,\forall i} r(i,t)^{\alpha} \\+
  \frac{1}{2}\sum_{\substack{i,\alpha,\\ j,\beta}}\left.\frac{\partial^2V}
{\partial r(i,t)^{\alpha}\partial r(j,t)^{\beta}}\right|_{\vec{r}(i,t)=0,\forall i}
  r(i,t)^{\alpha}r(j,t)^{\beta}\\+
\ldots,
\label{eq:PE_harmonic}
\end{multline}
where $r(i,t)^{\alpha}$ is the $\alpha$-cartesian component ($\alpha=x,y,z$) of the time-dependent atomic displacement
$\vec{r}(t)$ of the $i^{\mathrm{th}}$ atom about its equilibrium position,
$\left.V\right|_{\vec{r}(i,t)=0,\forall i}$ is the potential energy of the crystal in its equilibrium configuration,
$\left.\partial V/\partial r(i,t)^{\alpha}\right|_{\vec{r}(i,t)=0,\forall i}$
is the negative of the force acting in the $\alpha$ direction on atom $i$ in the equilibrium configuration
(zero by definition), and
$\left.\partial^2V/\partial r(i,t)^{\alpha}\partial r(j,t)^{\beta}\right|_{\vec{r}(i,t)=0,\forall i}$
constitute the \IFC{}s $\phi(i,j)_{\alpha,\beta}$.
To first approximation, $\phi(i,j)_{\alpha,\beta}$ is the negative of the force exerted
in the $\alpha$ direction on atom $i$ when atom
$j$ is displaced in the $\beta$ direction with all other atoms maintaining their equilibrium positions,
as shown in Figure
\ref{fig:thermomechanical}(c).
All higher order terms are neglected in the harmonic approximation.

Correspondingly, the equations of motion of the lattice are as follows:
\begin{equation}
  M(i)\ddot{r}(i,t)^{\alpha}=
-\sum_{j,\beta}\phi(i,j)_{\alpha,\beta}
r(j,t)^{\beta}\quad\forall i,\alpha,
\label{eq:eom_harmonic}
\end{equation}
and can be solved by a plane wave solution of the form
\begin{equation}
r(i,t)^{\alpha}=\frac{v(i)^{\alpha}}{\sqrt{M(i)}} e^{\mathrm{i}\left(\vec{q}\cdot\vec{R}_{l} - \omega t \right)},
\end{equation}
where $v(i)^{\alpha}$ form the phonon eigenvectors (polarization vector),
$M(i)$ is the mass of the $i^{\mathrm{th}}$ atom,
$\vec{q}$ is the wave vector,
$\vec{R}_{l}$ is the position of lattice point $l$,
and $\omega$ form the phonon eigenvalues (frequencies).
The approach is nearly identical to that taken for electrons in a periodic potential (Bloch waves) \cite{ashcroft_mermin}.
Plugging this solution into the equations of motion (Equation~\ref{eq:eom_harmonic}) yields the following set of linear equations:
\begin{equation}
\omega^{2}v(i)^{\alpha}=
  \sum_{j,\beta}D_{i,j}^{\alpha,\beta}(\vec{q})
  v(j)^{\beta}\quad\forall i,\alpha,
\end{equation}
where the dynamical matrix $D_{i,j}^{\alpha,\beta}(\vec{q})$ is defined as
\begin{equation}
D_{i,j}^{\alpha,\beta}(\vec{q})=
  \sum_{l}\frac{\phi(i,j)_{\alpha,\beta}}{\sqrt{M(i)M(j)}} e^{-\mathrm{i}\vec{q}\cdot\left(\vec{R}_{l}-\vec{R}_{0}\right)}.
\end{equation}
The problem can be equivalently represented by a standard eigenvalue equation:
\begin{equation}
\omega^{2}
\begin{bmatrix}
                                            \\
\vec{v} \\
 ~
\end{bmatrix}
=
\begin{bmatrix}
 &                                                                                       & \\
  & \mathbf{D}(\vec{q}) & \\
 &                                                                                       &
\end{bmatrix}
\begin{bmatrix}
                                            \\
\vec{v} \\
 ~
\end{bmatrix},
\label{eq:dyn_eigen}
\end{equation}
where
the dynamical matrix and phonon eigenvectors have dimensions $\left(3 n_{\mathrm{a}} \times 3 n_{\mathrm{a}}\right)$
and $\left(3 n_{\mathrm{a}} \times 1 \right)$, respectively, and $n_{\mathrm{a}}$ is the number of atoms in the cell.
Hence, Equation \ref{eq:dyn_eigen} has $\lambda=3 n_{\mathrm{a}}$ solutions/modes referred to as branches.
In practice, Equation \ref{eq:dyn_eigen} is solved for discrete sets of $\vec{q}$-points to compute
the phonon density of states (grid over all possible $\vec{q}$) and dispersion
(along the high-symmetry paths of the lattice \cite{curtarolo:art58}).
Thus, the phonon eigenvalues and eigenvectors are appropriately denoted $\omega_{\lambda}(\vec{q})$ and
$\vec{v}_{\lambda}(\vec{q})$, respectively.

Similar to the electronic Hamiltonian, the dynamical matrix is Hermitian, \textit{i.e.},
$\mathbf{D}(\vec{q})=\mathbf{D}^{*}(\vec{q})$.
Thus $\omega_{\lambda}^{2}(\vec{q})$ must also be real, so $\omega_{\lambda}(\vec{q})$ can either be real or purely imaginary.
However, a purely imaginary frequency corresponds to vibrational motion of the lattice that increases exponentially in time.
Therefore, imaginary frequencies, or those corresponding to soft modes, indicate the structure is dynamically unstable.
In the case of a symmetric, high-temperature phase, soft modes suggest there exists a lower symmetry structure
stable at $T=0$~K.
Temperature effects on phonon frequencies can be modeled with
\begin{equation}
\widetilde{\omega}_{\lambda}^{2}(\vec{q},T)=\omega_{\lambda}^{2}(\vec{q},T=0)+\eta T^2,
\end{equation}
where $\eta$ is positive in general.
The two structures, the symmetric and the stable, differ by the distortion
corresponding to this ``frozen'' (non-vibrating) mode.
Upon heating, the temperature term increases until the frequency reaches zero, and a phase transition occurs from
the stable structure to the symmetric~\cite{Dove_LatDynam_1993}.

In practice, soft modes~\cite{Parlinski_Phonon_Software} may indicate:
\textbf{i.} the structure is dynamically unstable at $T$,
\textbf{ii.} the symmetry of the structure is lower than that considered, perhaps due to magnetism,
\textbf{iii.} strong electronic correlations, or
\textbf{iv.} long range interactions play a significant role, and a larger supercell should be considered.

With the phonon density of states computed, the following thermal properties can be calculated:
the internal vibrational energy
\begin{equation}
  U_\svib(\vec{x},T)=\int_{0}^{\infty} \left( \frac{1}{2} + \frac{1}{e^{\left(\beta \hbar \omega\right)}-1} \right) \hbar \omega g(\vec{x};\omega) d\omega,
\end{equation}
the vibrational component of the free energy $F_\svib(\vec{x}; T)$ (Equation~\ref{eq:fvib}),
the vibration entropy
\begin{equation}
S_\svib(\vec{x},T)=\frac{U_\svib(\vec{x},T)-F_\svib(\vec{x}; T)}{T},
\end{equation}
and the isochoric specific heat
\begin{equation}
C_{\sV, \svib}(\vec{x},T)=\int_{0}^{\infty} \frac{  k_\sB \left(\beta \hbar \omega \right)^2 g(\vec{x};\omega)}{\left(1-e^{-\left(\beta \hbar \omega\right)}\right) \left(e^{\left(\beta \hbar \omega\right)}-1\right) } d\omega.
\end{equation}

\subsection{Quasi-harmonic phonons}

The harmonic approximation does not describe phonon-phonon scattering, and so cannot be used to
calculate properties such as thermal conductivity or thermal expansion.
To obtain these properties, either the quasi-harmonic approximation can be used,
or a full calculation of the higher order anharmonic \IFC{}s can be performed.
The quasi-harmonic approximation is the less computationally demanding of these two methods,
and compares harmonic calculations of phonon properties at different volumes to predict anharmonic properties.
The different volume calculations can be in the form of harmonic phonon calculations as described
above \cite{curtarolo:art114, curtarolo:art119},
or simple static primitive cell calculations \cite{Blanco_CPC_GIBBS_2004, curtarolo:art96}.
The \underline{Q}uasi-\underline{H}armonic \underline{A}pproximation
is implemented within \APL\ and referred to as \QHAAPL~\cite{curtarolo:art96}.
In the case of the quasi-harmonic phonon calculations, the anharmonicity of the system is described by
the mode-resolved Gr{\"u}neisen parameters, which are given by the change in the phonon frequencies as a function of volume
\begin{equation}
  \label{gamma_micro}
  \gamma_{\lambda}(\vec{q}) = - \frac{V}{\omega_{\lambda}(\vec{q})} \frac{\partial \omega_{\lambda}(\vec{q})}{\partial V},
\end{equation}
where $\gamma_{\lambda}(\vec{q})$ is the parameter for the wave vector $\vec{q}$ and the $\lambda^{\rm{th}}$ mode of the phonon dispersion.
The average of the $\gamma_{\lambda}(\vec{q})$ values, weighted by the specific heat capacity of each mode $C_{\sV,\lambda}(\vec{q})$, gives the average
Gruneisen parameter:
\begin{equation}
  \label{gamma_ave}
  \gamma = \frac{\sum_{\lambda,\vec{q}} \gamma_{\lambda}(\vec{q})  C_{\sV,\lambda}(\vec{q})}{C_\sV}.
\end{equation}
The specific heat capacity, Debye temperature and Gr{\"u}neisen parameter can then be combined to
calculate other properties such as the specific heat capacity at constant pressure $C_{\rm p}$,
the thermal coefficient of expansion $\alpha$, and the lattice thermal
conductivity $\kappa_\sL$ \cite{curtarolo:art119}, using similar expressions to those described in Section \ref{sect:qha_debye}.

\subsection{Anharmonic phonons}

The full calculation of the anharmonic \IFC{}s requires performing supercell calculations in which pairs of
inequivalent atoms are displaced in all pairs of
inequivalent directions \cite{Broido2007, Wu_PRB_2012, ward_ab_2009, ward_intrinsic_2010, Zhang_JACS_2012, Li_PRB_2012, Lindsay_PRL_2013, Lindsay_PRB_2013, Li_ShengBTE_CPC_2014, curtarolo:art125}
as illustrated in Figure \ref{fig:thermomechanical}(c).
The third order anharmonic \IFC{}s can then be obtained by calculating the change in the forces on
all of the other atoms due to these displacements.
This method has been implemented in the form of a fully
automated integrated workflow in the \AFLOW\ framework,
where it is referred to as the {\small \underline{A}FLOW} \underline{A}nharmonic
\underline{P}honon \underline{L}ibrary (\AAPL) \cite{curtarolo:art125}.
This approach can provide very accurate results for the lattice thermal conductivity when combined
with accurate electronic structure methods
\cite{curtarolo:art125},
but quickly becomes very expensive for systems with multiple
inequivalent atoms or low symmetry.
Therefore, simpler methods such as the quasi-harmonic Debye model
tend to be used for initial rapid screening \cite{curtarolo:art96, curtarolo:art115}, while
the more accurate and expensive methods are used for characterizing systems
that are promising candidates for specific engineering applications.

\section{Online data repositories}

Rendering the massive quantities of data generated using automated \textit{ab-initio} frameworks available
for other researchers requires going beyond the conventional methods
for the dissemination of scientific results in the form of journal articles.
Instead, this data is typically made available in online data repositories, which can usually be accessed both
manually via interactive web portals, and programmatically via an \underline{a}pplication \underline{p}rogramming \underline{i}nterface (\API).

\subsection{Computational materials data web portals}

\begin{figure*}[t!]
\includegraphics[width=1.00\linewidth]{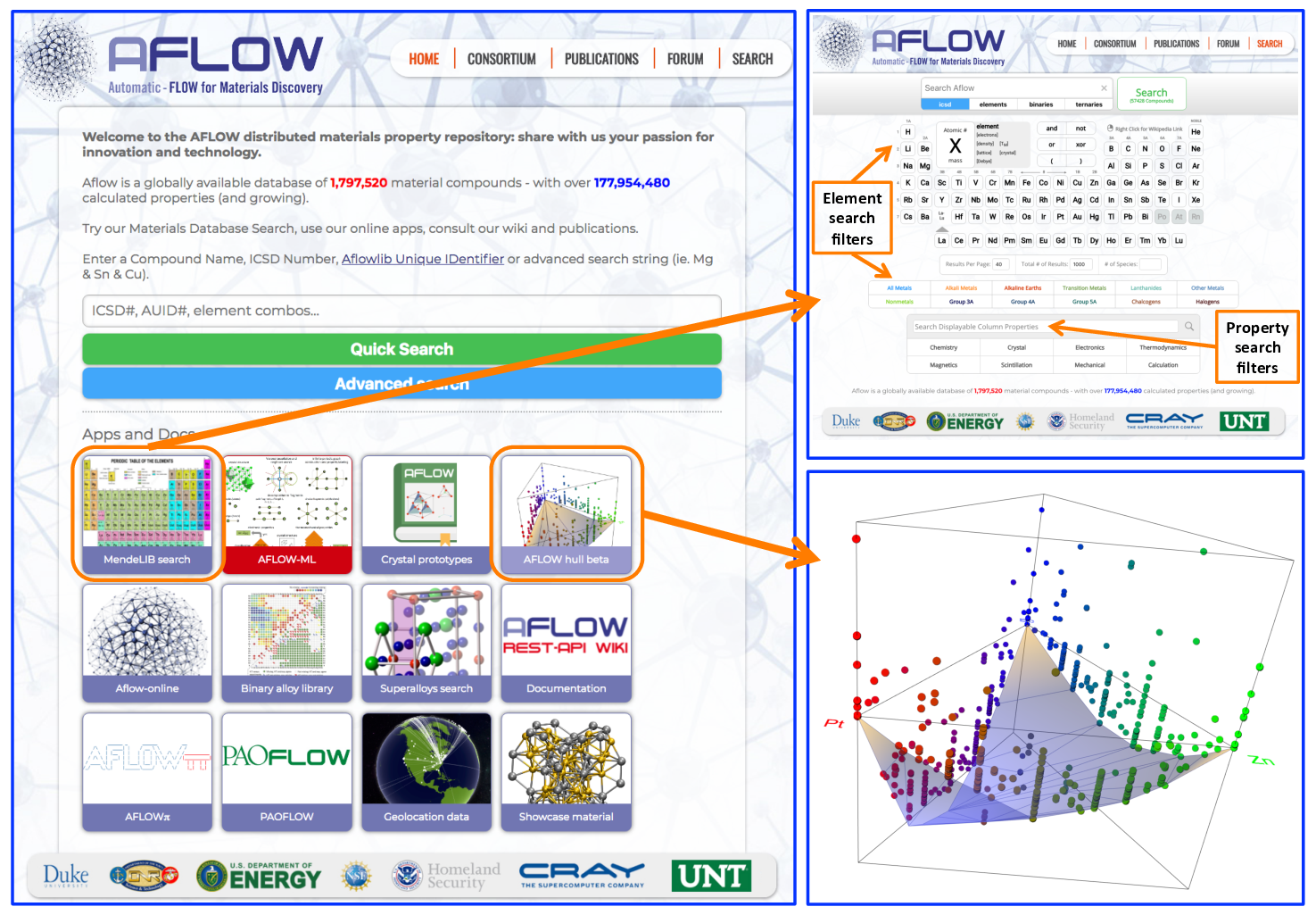}
\vspace{-4mm}
\caption{
\small
{\bf (a)} Front page of the \AFLOW\ online data repository, highlighting the link to
{\bf (b)} the \AFLOW\ advanced search application, which facilitates complex search
queries including filtering by chemical composition and materials properties and
{\bf (c)} the \AFLOW\ interactive convex hull generator, showing the 3D hull for the Pt-Sc-Zn ternary alloy system.
}
\label{fig:aflow_web_apps}
\end{figure*}

Most computational data repositories include an interactive web portal front end that enables manual data access.
These web portals usually include online applications to facilitate data retrieval and analysis.
The front page of the \AFLOW\ data repository is displayed in Figure \ref{fig:aflow_web_apps}(a).
The main features include a search bar where information such as
ICSD reference number, {\small \underline{A}FLOW} \underline{u}nique \underline{id}entifier (AUID) or the chemical formula can be entered
in order to retrieve specific materials entries.
Below are buttons linking to several different online applications such as the advanced search functionality,
convex hull phase diagram generators, machine learning applications \cite{curtarolo:art124, curtarolo:art129, curtarolo:art136} and \AFLOW-online data analysis tools.
The link to the advanced search application is highlighted by the orange square, and the application page is shown in Figure \ref{fig:aflow_web_apps}(b).
The advanced search application allows users to search for materials that contain (or exclude) specific elements or groups of elements,
and also to filter and sort the results by properties such as electronic band structure energy gap (under the ``Electronics'' properties filter group)
and bulk modulus (under the ``Mechanical'' properties filter group).
This allows users to identify candidate materials with suitable materials properties for specific applications.

Another example online application available on the \AFLOW\ web portal is the convex hull phase diagram generator.
This application can be accessed by clicking on the button highlighted by the orange square in
Figure \ref{fig:aflow_web_apps}(a), which will bring up a periodic table allowing users to
select two or three elements for which they want to generate a convex hull.
The application will then access the formation enthalpies and stoichiometries of the materials entries in the
relevant alloy systems, and use this data to generate a two or three dimensional convex hull phase diagram
as depicted in Figure \ref{fig:aflow_web_apps}(c).
This application is fully interactive, allowing users to adjust the energy axis scale,
rotate the diagram to view from different directions, and select specific points to obtain more information on the
corresponding entries.

\subsection{Programmatically accessible online repositories of computed materials properties}

In order to use materials data in machine learning algorithms, it should be stored in a structured online database
and made programmatically accessible via a \underline{re}presentational \underline{s}tate \underline{t}ransfer \API\ (\RESTAPI).
Examples of online repositories of materials data include \AFLOW\ \cite{aflowlibPAPER, curtarolo:art92},
Materials Project \cite{materialsproject.org}, and OQMD \cite{Saal_JOM_2013}.
There are also repositories that aggregate results from multiple sources such as
NoMaD \cite{nomad} and Citrine \cite{citrine_database}.

\begin{figure}[t!]
  \includegraphics[width=1.00\linewidth]{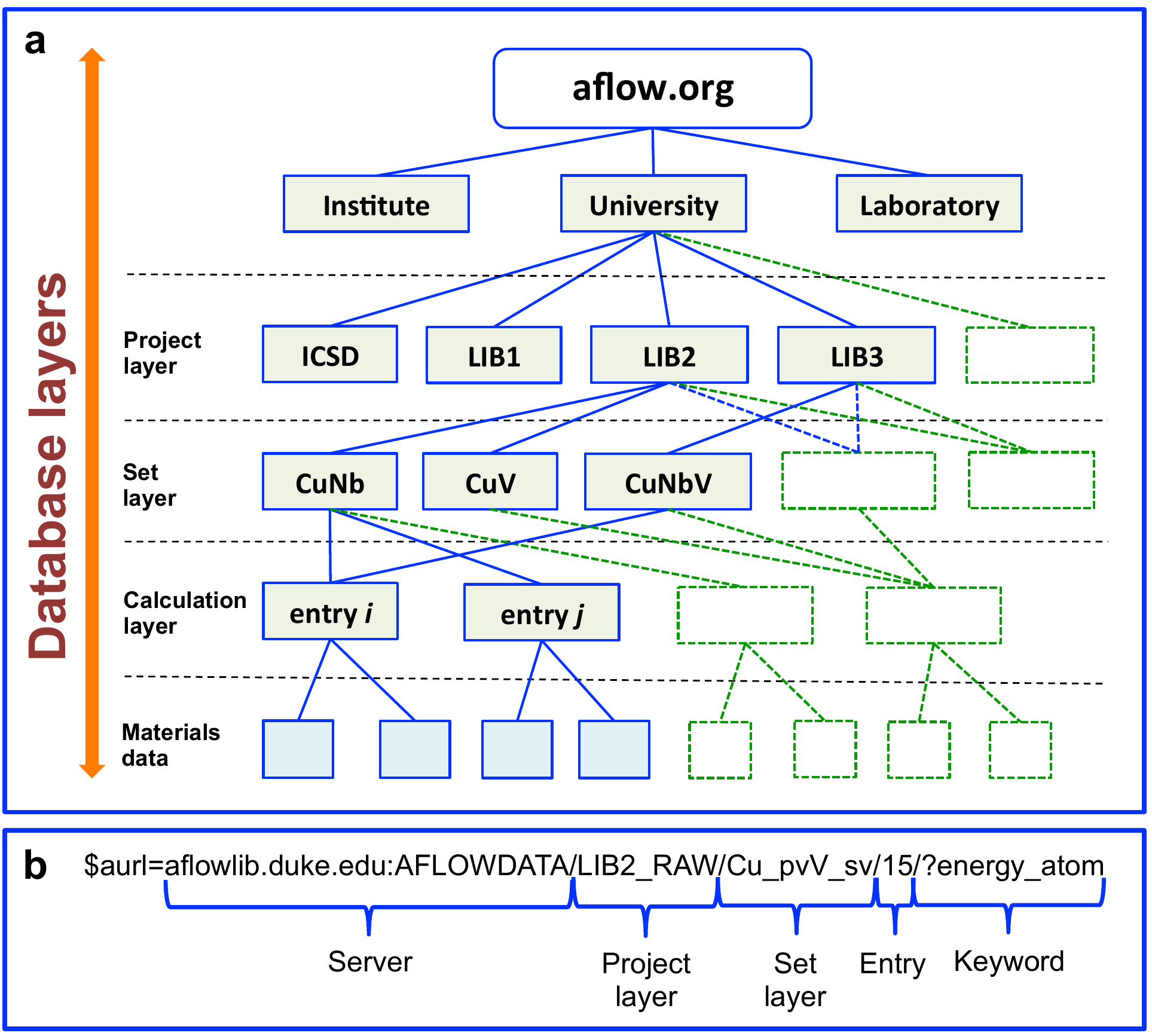}
  \vspace{-4mm}
  \caption{\small (a) The \AFLOW\ database is organized as a multilayered system.
(b) Example of an AURL which enables direct programmatic access to specific materials entry properties in the \AFLOW\ database.}
  \label{fig:aflow_restapi_layers_aurl}
\end{figure}

\RESTAPI{}s facilitate programmatic access to data repositories.
Typical databases such as \AFLOW\ are organized in layers,
with the top layer corresponding to a project or catalog (\textit{e.g.}, binary alloys),
the next layer corresponding to data sets (\textit{e.g.}, all of the entries for a particular alloy system),
and then the bottom layer corresponding to specific materials entries, as illustrated in Figure \ref{fig:aflow_restapi_layers_aurl}(a).

In the case of the \AFLOW\ database, there are currently four different ``projects'', namely the
``ICSD'', ``LIB1'', ``LIB2'' and ``LIB3'' projects; along with three
more under construction: ``LIB4'', ``LIB5'' and ``LIB6''.
The ``ICSD'' project contains calculated data for previously observed compounds \cite{ICSD},
whereas the other three projects contain calculated data for single elements, binary alloys,
and ternary alloys respectively, and are constructed by decorating prototype
structures with combinations of different elements.
Within ``LIB2'' and ``LIB3'', there are many different data sets, each corresponding to a specific
binary or ternary alloy system.
Each entry in the set corresponds to a specific prototype structure and stoichiometry.
The materials properties values for each of these entries are encoded via keywords,
and the data can be accessed via \URL{}s constructed from the different layer names and the appropriate keywords.
In the case of the \AFLOW\ database, the location of each layer and entry is
identified by an {\small \underline{A}FLOW} \underline{u}niform \underline{r}esource \underline{l}ocator (\AURL) \cite{curtarolo:art92},
which can be converted to a \URL\ providing the absolute path to a particular layer, entry or property.
The \AURL\ takes the form \url{server:AFLOWDATA/project/set/entry/?keywords},
for example \url{aflowlib.duke.edu:AFLOWDATA/LIB2_RAW/Cu_pvV_sv/15/?energy_atom},
where \url{aflowlib.duke.edu} is the web address of the physical server where the data is located,
\url{LIB2_RAW} is the binary alloy project layer, \url{Cu_pvV_sv} is
the set containing the binary alloy system Cu-V, \url{15} is a specific entry with the composition
Cu$_3$V in a tetragonal lattice, and \url{energy_atom} is the keyword corresponding
to the property of energy per atom in units of eV, as shown in Figure \ref{fig:aflow_restapi_layers_aurl}(b).
Each \AURL\ can be converted to a web \URL\ by changing the ``\url{:}'' after the server name to a ``\url{/}'',
so that the \AURL\ in Figure \ref{fig:aflow_restapi_layers_aurl}(b) would become the
\URL\ \url{aflowlib.duke.edu/AFLOWDATA/LIB2_RAW/Cu_pvV_sv/15/?energy_atom}.
This \URL, if queried via a web browser or using a UNIX utility such as \texttt{wget},
returns the energy per atom in eV for entry \url{15} of the Cu-V binary alloy system.

In addition to the \AURL, each entry in the \AFLOW\ database is also associated with an
\AUID\ \cite{curtarolo:art92},
which is a unique hexadecimal (base 16) number constructed from a checksum of the \AFLOW\ output file for that entry.
Since the \AUID\ for a particular entry can always be reconstructed by applying the checksum
procedure to the output file, it serves as a permanent, unique specifier for each calculation,
irrespective of the current physical location of where the data are stored.
This enables the retrieval of the results for a particular calculation from different servers,
allowing for the construction of a truly distributed database that is robust
against the failure or relocation of the physical hardware. Actual database versions can be
identified from the version of \AFLOW\ used to parse the calculation output files and
postprocess the results to generate the database entry. This information can be retrieved using the
keyword \url{aflowlib_version}.

The search and sort functions of the front-end portals can be combined with the programmatic
data access functionality of the \RESTAPI\ through the implementation of a Search-\API.
The \AFLUX\ Search-\API\ uses the \LUX\ language to enable the embedding of logical
operators within \URL\ query strings \cite{curtarolo:art128}.
For example, the energy per atom of every entry in the \AFLOW\ repository containing the element Cu or V,
but not the element Ti, with an electronic band gap between 2 and 5~eV, can be retrieved using the command:
\url{aflowlib.duke.edu/search/API/species((Cu:V),(!Ti)),Egap(2*,*5),energy_atom}.
In this \AFLUX\ search query, the comma ``\verb|,|'' represents the logical {\small AND} operation, the colon ``\verb|:|'' the logical {\small OR} operation,
the exclamation mark ``\verb|!|'' the logical {\small NOT} operation, and the asterisk ``\verb|*|''  is the ``loose'' operation that defines a range of values to search within.
Note that by default \AFLUX\  returns only the first 64 entries matching the search query.
The number and set of entries can be controlled by appending the \verb|paging| directive to the end of the search query as follows:
\url{aflowlib.duke.edu/search/API/species((Cu:V),(!Ti)),Egap(2*,*5),energy_atom,paging(0)},
where calling the  \verb|paging| directive with the argument ``0'' instructs \AFLUX\ to return all of the matching entries
(note that this could potentially be a large amount of data, depending on the search query).
The \AFLUX\ Search-\API\ allows users to construct and retrieve customized data sets, which they can feed into materials
informatics machine learning packages to identify trends and correlations for use in rational materials design.

The use of \API{}s to provide programmatic access is being extended beyond materials data retrieval,
to enable the remote use of pre-trained machine learning algorithms.
The \AFLOWML\ \API\ \cite{curtarolo:art136} facilitates access to the two machine learning models
that are also available online at \url{aflow.org/aflow-ml} \cite{curtarolo:art124, curtarolo:art129}.
The \API\ allows users to submit structural data for the material of interest using a utility such as \verb|cURL|,
and then returns the results of the model's predictions in \JSON\ format.
The programmatic access to machine learning predictions enables the incorporation of machine learning into
materials design workflows, allowing for rapid pre-screening to automatically select
promising candidates for further investigation.

\section{Materials applications}

The automated approach to computational materials science has been used to accelerate the design of materials for structural applications
such as metallic glasses and superalloys, and for functional applications including thermoelectrics, magnets, catalysts,
batteries, photovoltaics and superconductors.

\subsection{Disordered Materials}

Section \ref{htframeworks} describes how the thermodynamic stability of ordered compounds at zero
temperature can be predicted from the convex hull phase diagrams generated using the
formation enthalpies available in computational materials data repositories such as
\AFLOW\ \cite{curtarolo:art65, aflowlibPAPER, curtarolo:art92, curtarolo:art104}.
At finite temperature, however, entropic contributions
due to thermally driven disorder play an important role, and lead to the formation of
disordered materials such as metallic glasses and solid solutions.
The thermodynamically favored phase at a given temperature and pressure is the phase with the lowest Gibbs free energy.
Since the entropy term in the Gibbs free energy is multiplied by the temperature $T$,
the entropic contribution to the Gibbs free energy becomes increasingly important at higher temperatures.
The entropy of materials has two main components: the vibrational entropy, $S_\svib$, which can be calculated from the phonon
dispersion or the Debye model as described in Section \ref{sect:thermomechanical};
 and the configuration entropy, $S_\sconfig$, due to the disorder in the atomic positions or site occupations.
Configurational entropy originates from chemical disorder as in the case of high entropy alloys in which all of the atoms are arranged on a regular lattice,
but the specific lattice sites are randomly occupied by different chemical species;
or structural disorder as in the case of metallic glasses, where the atoms no longer occupy regular lattice sites resulting in an amorphous material.

\noindent
\textbf{High entropy materials.}
High entropy materials display structural order (\textit{i.e.} all of the atoms are arranged on a periodic crystal lattice)
but chemical disorder (\textit{i.e.} the actual occupation of these lattice sites is random) \cite{widom2015high}.

In the ideal entropy limit in which the occupation of the atomic sites is completely random,
the configuration entropy per atom is given by $S_\sconfig = k_\sB \sum_i x_i \log_e \left( x_i \right)$ \cite{widom2015high},
where $x_i$ is the fractional composition of each species component.
Note that this expression increases with increasing numbers of species, and is also maximized when all
of the values of $x_i$ are equal, \textit{i.e.} for equimolar compositions.

The expression for the ideal entropy can be combined with calculations for
\underline{s}pecial \underline{q}uasirandom \underline{s}tructures (\SQS) \cite{zunger_sqs},
which are special structural configurations where the radial
correlation functions mimic those of a perfectly random structure, to estimate the
Gibbs free energy for high entropy alloys.
This can then be used in conjunction with the energies of the ordered
phases obtained from computational materials data repositories such
as \AFLOW\ \cite{curtarolo:art65, aflowlibPAPER, curtarolo:art92, curtarolo:art104}
to generate structural phase diagrams as a function of temperature and composition,
predicting the phase transition boundaries between ordered compounds, phase separation regions,
and single phase solid solutions \cite{curtarolo:art106, curtarolo:art117, curtarolo:art126}.
The calculated ordered structure energies in \AFLOW\ can also be used to train cluster expansion models \cite{Walle_calphad_2002} to predict
the energies of large ensembles of configurations, which can be combined with thermodynamic
descriptors to estimate the transition temperature and miscibility gaps for solid solutions and
high entropy alloys \cite{Lederer_HEA_2018}.

The concept of entropy stabilization has recently been extended beyond metallic alloys to
include multi-component ceramics, such as high entropy oxides
\cite{curtarolo:art99, curtarolo:art122}.
High entropy oxides consist of an ordered anion sublattice occupied by oxygen ions,
with a disordered cation sublattice randomly occupied by five different metal ions,
such as Co, Cu, Mg, Ni and Zn \cite{curtarolo:art99, curtarolo:art122}.
The oxygen ions screen the metal ions from each other, reducing the energy cost associated with
forming a random configuration of the metal ions, enabling the formation of a single-phase, entropy-stabilized ceramic.

\noindent
\textbf{Metallic glasses.}
Metallic glasses are alloys in which the atoms do not occupy the sites of a
regular periodic lattice, but instead form a structurally disordered amorphous phase.
These materials are of great commercial and industrial interest due to their
unique combination of superb mechanical properties \cite{chen2015does} and
plastic-like processability \cite{schroers2006amorphous, Schroers_blow_molding_2011,
kaltenboeck2016shaping} for several potential applications \cite{Johnson_BMG_2009,
Greer_metallic_glasses_review_2009, Schroers_Processing_BMG_2010, johnson2016quantifying, ashby2006metallic}.

Several different attempts have been made to understand the formation of metallic glasses and predict the
\underline{g}lass \underline{f}orming \underline{a}bility (\GFA) of different alloy compositions.
Most of these efforts center around maximizing the packing density of the different atoms \cite{miracle2004structural},
which requires elements with a range of different atomic radii \cite{egami1984atomic, greer1993confusion, egami2003atomistic, lee2003criteria, zhang2015origin}.
Other efforts have been made to use phase diagram data on liquidus temperatures to predict \GFA\ \cite{cheney2009evaluation, cheney2007prediction, lu2002new}.
Work is also underway to use machine learning techniques to predict potential glass formers \cite{Ward_ML_GFA_NPGCompMat_2016}.

Much of the theoretical work described above relies on the
use of experimental rather than \textit{ab-initio} computational data to predict new materials, due to the difficulty
of modeling amorphous structures using first-principles techniques.
However, Perim \textit{et al.} \cite{curtarolo:art112} recently demonstrated that the energies of different structural phases can be
combined into a descriptor to predict the formation of metallic glasses.
If there are many different structural phases with similar formation enthalpy,
this will frustrate crystallization during solidification and thus promote
glass formation \cite{curtarolo:art112}.
This frustration can be quantified to formulate a spectral descriptor for \GFA\ using the structural and energetic information available in computational
materials data repositories such as \AFLOW\ \cite{curtarolo:art65, aflowlibPAPER, curtarolo:art92, curtarolo:art104}.
The differences in the geometry between two structures are quantified by describing
each structure in terms of its atomic environments \cite{villars:factors, daams:environments_book, daams_villars:environments_2000},
while the formation enthalpy differences between the respective structures are expressed in the form of Boltzmann factors.
The energetic and structural descriptors can then be combined with appropriate normalization
factors to formulate a spectral descriptor for \GFA\ as function of composition $x$: $GFA \left(\{x\}\right)$.
Comparisons with known glass forming compositions available in the literature can then be used to define a threshold,
such that if $GFA \left(\{x\}\right)$ exceeds this threshold, then the composition $x$ would be expected to be glass forming.

The $GFA \left(\{x\}\right)$ descriptor has been used to perform an automated analysis of the \GFA\
of over 1,400 binary alloy systems from the \AFLOW\ data repository \cite{curtarolo:art112}.
While over half of all binary alloy systems are predicted to have a \GFA\ below that of the threshold,
nevertheless some 17\% of alloy systems display a maximum value of $GFA \left(\{x\}\right)$
greater than the maximum value for the Cu-Zr system, a well-known good glass former.
These included several alloy systems for which glass formation had never previously been observed
or sometimes even investigated, suggesting that there are many possible glass
forming compositions which remain to be discovered.
This success demonstrates the power of combining descriptors based on the easily
calculated properties of periodic crystalline phases with large pre-calculated databases for predicting
the synthesizability of complex disordered materials.

\noindent
\textbf{Modeling off-stoichiometry materials.}
Incorporating the effects of disorder is a necessary, albeit difficult, step in materials modeling.
Not only is disorder intrinsic to all materials,
but it also offers a route to enhanced and even otherwise inaccessible functionality,
as demonstrated by its ubiquity in technological applications.
Prominent examples include fuel cells~\cite{Xie_ACatB_2015},
high-temperature superconductors~\cite{high_Tc,Maeno_Nature_1994},
and low thermal conductivity thermoelectrics~\cite{Winter_JACerS_2007}.

Specifically, chemical disorder can arise in the form of doping, vacancies, and even
in the occupation of lattice sites themselves (random), which cannot inherently be
modeled using periodic systems.
One approach for modeling such effects includes \SQS~\cite{zunger_sqs}.
These quasirandom approximates are very computationally effective, but only offer
a single representation of the disordered states, \textit{i.e.}, that with the lowest site correlations.
Instead of reducing down to a single representation,
\AFLOW\ treats such systems as an ensemble of ordered supercells~\cite{curtarolo:art110}.
Properties are resolved through ensemble averages of the representative states,
with opportunities to optimize computation (via supercell size/site error) and
tune the level of disorder explored (via parameter $T$).
\AFLOWPOCC\ (\AFLOW\ \underline{p}artial \underline{occ}upation module) has already resolved
significant stoichiometric trends in wide-gap semiconductors and magnetic systems
while offering additional insight into underlying physical mechanisms.
Ultimately, the screening criteria and property predictions generated by these \textit{bona fide} thermodynamic models
and descriptors are accelerating design of new, technologically-significant materials,
including advanced ceramics \cite{curtarolo:art80} and metallic glasses \cite{curtarolo:art112}.

\subsection{Superalloys}

Superalloys are characterized by their extraordinary mechanical properties, particularly at temperatures near their melting point.
Such traits make them the ideal candidates for applications in the aerospace and power generation industries.
Among the more common examples, many have a face-centered cubic structure with base elements nickel, cobalt, and iron, though
nickel-based superalloys dominate the market.
Surprisingly, a novel cobalt-based superalloy, Co$_{3}$(Al,W) was discovered in 2006 that exhibits mechanical properties better than many
nickel-based superalloys.
This inspired a thorough computational investigation with \AFLOW\ of alloys containing 40 different elements, yielding over 2,224 relevant ternary systems~\cite{curtarolo:art113}.
The search offered 102 systems shown to \textbf{i.} be more stable than Co$_{3}$[Al$_{0.5}$,W$_{0.5}$] --- the L1$_{2}$-like random structure
previously characterized thermodynamically~\cite{Saal_ActMat_2013} and very close to the compositions reported by experiments~\cite{Sato_Science_2006},
\textbf{ii.} have a relevant concentration $\left(X_{3}[A_{x}B_{1-x}]\right)$ that is in two-phase equilibrium with the host matrix, and
\textbf{iii.} exhibit only small deviations from the host matrix lattice (within 5\% relative mismatch).

For these 102 candidates, additional pertinent properties were extracted, including the density and bulk modulus (as a proxy for hardness).
Low density materials are preferred to mitigate the stress on turbine components.
Significant trends for the bulk modulus are elucidated when plotting with respect to component $B$ on a Pettifor scale:
Ni-based materials show a peak at or before Ni, whereas Co-based materials monotonically increase.
Additionally, Co-based materials are generally more resistant to compression compared to Ni-based materials.

Of the 102 candidates, 37 materials have no reported phase diagrams in standard databases, and are thus expected
to be unexplored or new.
Additional screening based on the toxicity and (low) melting temperature of components uncovered
six priority candidates for experimental validation.

\subsection{Thermoelectrics}

Thermoelectric materials generate an electric voltage when subjected to a temperature gradient,
and can also generate a temperature gradient when a voltage is applied
\cite{snyder_complex_2008, nolas_thermoelectrics:_2001}.
Their lack of moving parts and resulting scalability means that they have potential applications
in power generation for spacecraft, energy recovery from waste heat in automotive and industrial facilities \cite{bell_cooling_2008, disalvo99}
and in spot cooling for nanoelectronics using the Peltier cooling effect \cite{bell_cooling_2008, disalvo99}.
However, most of the available thermoelectric materials have low efficiency, only converting a few percent
of the available thermal energy into electricity. Therefore, a major goal of thermoelectrics research is to develop new materials that have
higher thermoelectric efficiency.

The thermoelectric efficiency of a material is determined by the figure of merit $z T$, which is obtained from \cite{snyder_complex_2008, nolas_thermoelectrics:_2001}
\begin{equation}
\label{thermopower}
z T = \frac{\sigma {\sf S}^2 T}{\kappa_\sL + \kappa_{\mathrm{e}}},
\end{equation}
where {\sf S} is the Seebeck coefficient, $\sigma$ is the electrical conductivity,
$\kappa_\sL$ is the lattice thermal conductivity, and $\kappa_{\mathrm{e}}$ is
the electronic thermal conductivity.
The lattice thermal conductivity $\kappa_\sL$ can be calculated using
the methods described in Section \ref{sect:thermomechanical}.
Most of the electronic thermal conductivity $\kappa_{\mathrm{e}}$ will depend directly on the
electrical conductivity $\sigma$ through the Wiedemann-Franz law \cite{snyder_complex_2008}
\begin{equation}
\label{electronic_kappa}
\kappa_{\mathrm{e}} = L \sigma T,
\end{equation}
where $L$ is the Lorenz factor, which has a value of $2.4 \times 10^{-8}$ J$^2$ K$^{-2}$ C$^{-2}$ for free electrons.
The Seebeck coefficient {\sf S} is given by \cite{snyder_complex_2008}
\begin{equation}
\label{Seebeck}
{\sf S} = \frac{8 \pi^2 k_\sB^2}{3 e h^2} m^* T \left(\frac{\pi}{3n} \right)^{\frac{2}{3}},
\end{equation}
where $n$ is the charge carrier concentration, $e$ is the electronic charge, and $m^*$ is the density of states effective mass of the charge carriers in the material.
The effective mass tensor $m_{ij}$ can be calculated from the curvature of electronic band structure dispersion $E(\vec{k})$
\begin{equation}
\label{mass_tensor}
m_{ij}^{-1} = \frac{1}{\hbar^2} \frac{d^2E}{dk_i dk_j},
\end{equation}
where $k_i$ and $k_j$ are components of the wave vector $\vec{k}$.
Larger curvature of the band structure implies a lower effective mass, while flat narrow bands tend to result in a large effective mass.
Charge carrier mobility and thus electrical conductivity tend to reduce with increasing effective mass.
However, as can be seen from Equation \ref{Seebeck}, the Seebeck coefficient increases with effective mass, and $\kappa_{\mathrm{e}}$ also increases with $\sigma$.
Therefore, a compromise should be found between high effective mass to maximize  {\sf S} and high charge carrier mobility to give high $\sigma$ in order to optimize the thermoelectric efficiency of the device.

Several computational high-throughput searches have been performed for
thermoelectric materials \cite{curtarolo:art68, madsen2006, curtarolo:art84, curtarolo:art85, Garrity_thermoelectrics_PRB_2016, Chen_thermoelectrics_JMCC_2016, Zhu_thermoelectrics_JMCC_2016, curtarolo:art105}.
Many of the efforts towards developing more efficient thermoelectric materials
have focused on either lowering the lattice thermal conductivity $\kappa_\sL$,
or finding materials in which the electronic properties
are highly directional, allowing for a narrow energy band distribution while simultaneously having
a low effective mass, thus increasing the power factor $\sigma {\sf S}^2$.
High-throughput searches for materials with low lattice thermal conductivity have
focused on materials such as half-Heusler structures \cite{curtarolo:art84, curtarolo:art85, Zeier_thermoelectrics_zintl_NRM_2016},
which have lower densities and thus lower thermal conductivities than the full Heusler structures.
Other promising materials include structures such as
clathrates \cite{Shi_clathrates_AFM_2010, Zhang_clathrates_IC_2011, Saiga_clathrates_JAC_2012, Christensen_clathrates_DT_2009, Madsen_PSSA_2016}
and skutterudites \cite{Sales_skutterudites_Sci_1996, Bai_skutterudites_AM_2009, curtarolo:art105, yang_trends_2011}, which contain hollow voids that
can be filled with ``rattler'' atoms to reduce the lattice thermal conductivity.
Filled skutterudites in particular, such as $R_x$Co$_4$Sb$_{12}$, are excellent thermoelectric materials because of their
combination of a high effective mass with high carrier mobility due to the existence of a secondary conduction band with
12 conducting charge carrier pockets \cite{curtarolo:art105}.

Searches of large databases of inorganic materials to find new thermoelectrics include the study of 48,000 materials from
the Materials Project database \cite{Chen_thermoelectrics_JMCC_2016},
where the power factor was calculated using the BoltzTraP code \cite{boltztrap}
and the thermal conductivity was estimated using the Clarke \cite{Clarke_thermal_conductivity_SCT_2003} and Cahill-Pohl \cite{Cahill_APR_2014} models.
Almost 600 oxides, nitrides and sulfides from the ICSD were investigated by Garrity \cite{Garrity_thermoelectrics_PRB_2016},
where the lattice thermal conductivity was calculated at the quasiharmonic phonon level of approximation, with particular attention being paid to degeneracies in the
conduction band minimum, or materials with strongly anisotropic conduction bands,
that result in an effective low-dimensional conductor with a corresponding increase in the power factor.
The thermoelectric material LiZnSb was proposed by an automated search of the calculated band structures of 1,640 compounds
in the ICSD containing Sb \cite{madsen2006}, although later experimental measurements did not
find a high thermoelectric efficiency for this compound \cite{toberer_thermoelectric_2009}.

Other strategies to increase the power factor include engineering the
band structure \cite{Pei_BandEngineering_AdvMat_2012} through volume changes by alloying different materials to create solid solutions,
such as antifluorite Mg$_2$Si and Mg$_2$Ge with Mg$_2$Sn, or orthorhombic Ca$_2$Si and Ca$_2$Ge with
Ca$_2$Sn \cite{Bhattacharya_thermoelectrics_PRB_2015}. Tuning the composition of alloys can also be used to converge the valence and conduction bands, enabling
high valley degeneracy to be achieved in materials such as PbTe$_{1-x}$Se$_x$ alloys \cite{Pei_BandConvergence_Nature_2011}. Solid solutions can also produce local anisotropic
structural disorder, increasing phonon scattering and thus improving the thermoelectric efficiency \cite{Zeier_Stannite_JACS_2012, Zeier_SolidSolution_JACS_2012}.

The exploitation of thermodynamic phenomena such as spinodal decomposition to self-assemble heterostructures with increased phonon scattering \cite{curtarolo:art107}
has also been proposed to enhance the efficiency of thermoelectric devices.
In this approach, materials such as PbSe and PbTe, that are miscible at high temperatures, undergo phase separation when the mixture is cooled slowly,
creating a layered heterostructure with a network of boundaries between the different components, which scatter phonons and thus suppress the thermal conductivity.
This concept has also been extended to other nanotechnology applications, \textit{e.g.} as a means to embed a network of electrically conducting nanowires, in the form of topologically
protected interface states, within an insulating matrix \cite{curtarolo:art134}.

The combination of different competing materials properties that must be optimized to maximize the thermoelectric efficiency highlights
the importance of integrated frameworks such as \AFLOW, which can automatically calculate different types of materials properties such as thermal
conductivity and electronic band structures. Having all of these electronic and thermal properties calculated and available in an integrated, searchable, sortable
data repository such as \AFLOWorg\ accelerates the design of new, high-efficiency thermoelectric materials.

\subsection{Magnetic materials}

The search for new magnetic systems remains a longstanding challenge
despite their ubiquity in modern technology~\cite{curtarolo:art109}.
Magnetism demonstrates remarkable sensitivity to a number of properties, including
electronic configuration, bond length/angle, and magnetic ion valence, and thus
its presence is rather uncommon and difficult to predict.
In fact, only two percent of the known inorganic compounds~\cite{ICSD} exhibit magnetic order of any kind.
Consumer applications place additional practical restrictions for magnets,
with the current global market effectively populated by only two dozen compounds.
These obstacles motivated a large-scale computational search with \AFLOW\ for new magnets among the Heusler structure family.
Heusler structures are of particular interest for a number of reasons:
\textbf{i.} several are known high-performance magnets,
\textbf{ii.} the breadth of distinct compounds offers an excellent chance for discovery,
\textbf{iii.} the full set of materials will likely offer other types of interesting materials
(aside from magnets), and
\textbf{iv.} they are metallic and thus well described by \DFT.
There are three types of Heusler structures, \textit{i.e.}, the regular- $X_{2}YZ$ (Cu$_{2}$MnAl-type),
inverse- $\left(XY\right)XZ$ (Hg$_{2}$CuTi-type), and the half-Heuslers $XYZ$ (MgCuSb-type).
By decorating these prototypes with ternary combinations of 55 elements, a total of 236,115 compounds were
generated and added to the \AFLOWorg\ repository.

As a first attempt, the analysis is limited to Heuslers containing elements of the
3$d$, 4$d$, and 5$d$ periods, \textit{i.e.}, a subset of 36,540 compounds.
Of this set, 248 are determined to be thermodynamically stable and 22
have a magnetic ground state compatible with the unit cells considered.
Among these 22 magnetic ground state compounds, a few prominent classes can be identified, including
Co$_{2}YZ$ and Mn$_{2}YZ$.
Upon further analysis of these classes, four materials became of particular interest.
In the first class Co$_{2}YZ$, there already exists 25 known compounds all
lying on the Slater-Pauling curve (magnetic moment per formula unit \textit{versus} number
of valence electrons)~\cite{Graf_PSSC_2011}.
The regression predicts Co$_{2}$MnTi to have the notably high Curie transition temperature
$T_{\mathrm{C}}$ of 940~K --- a feature shared by only two dozen known magnets.
The second class Mn$_{2}YZ$ is of interest because of their high $T_{\mathrm{C}}$
and potentially large magnetocrystalline anisotropy~\cite{Kreiner_ZAAC_2014}.
Two known examples from this class, Mn$_{2}$VAl and Mn$_{2}$VGa, show ferrimagnetic ordering,
matching two candidates from the list of 22, Mn$_{2}$PtCo and Mn$_{2}$PtV.
One more compound was highlighted for satisfying a stringent thermodynamic constraint.
Mn$_{2}$PdPt shows to be robustly stable by at least 30 meV, where the criterion
derives from the distance of the stable phase from the pseudo convex hull that neglects it.
This criterion quantifies the impact of the structure on the minimum energy surface.

Following an attempt to synthesis these four candidates, two were successful (Co$_{2}$MnTi
and Mn$_{2}$PtPd) and the other two decomposed into binary compounds.
In fact, Co$_{2}$MnTi shows a $T_{\mathrm{C}}$ of 938~K, almost exactly as predicted by
the Slater-Pauling curve.
Surprisingly, Mn$_{2}$PdPt shows antiferromagnetic ordering and tetragonal distortion
($c/a \sim 1.8$), a result corroborated by calculation upon further analysis.
Beyond the synthesis of these two systems, this investigation offers a new, accelerated pathway
to discovery over traditional trial-and-error approaches.

\section{Conclusion}

Automated computational materials design frameworks have the capability to rapidly generate materials data without the need for laborious human intervention.
They are being used to construct large repositories of programmatically accessible materials properties,
calculated in a standardized, consistent fashion so as to facilitate the identification of trends
and the training of machine learning models to predict electronic, thermal and mechanical behavior.
When combined with physical models and intelligently formulated descriptors, the data becomes a powerful tool
to accelerate the discovery of new materials for applications ranging from high-temperature
superalloys to thermoelectrics and magnets.

\section{Acknowledgments}
We thank Drs. S. Barzilai, Y. Lederer, O. Levy, F. Rose, P. Nath, D. Usanmaz, D. Hicks, E. Gossett, D. Ford, R. Friedrich,
M. Esters, P. Colinet, E. Perim, C. Calderon, K. Yang, M. Mehl, M. Buongiorno Nardelli, M. Fornari, G. Hart,  I. Takeuchi,
E. Zurek, P. Avery, R. Hanson, A. Kolmogorov, A. Natan, N. Mingo, J. Carrete, S. Sanvito, D. Brenner, K. Vecchio,
M. Scheffler, L. Ghiringhelli, O. Isayev, A. Tropsha, J. Schroers and  J. J. Vlassak for insightful discussions.
This work is supported by DOD-ONR (N00014-16-1-2326, N00014-16-1-2583, N00014-17-1-2090, N00014-17-1-2876),
by NSF (DMR-1436151), and by the Duke University---Center for Materials Genomics.
SC acknowledges support by the Alexander von Humboldt-Foundation for financial support.
CO acknowledges support from the NSF Graduate Research Fellowship \#DGF1106401.

\newcommand{\Ozolins}{Ozoli\c{n}\v{s}}

\end{document}